\documentclass[twocolumn,english,pra,aps,superscriptaddress,floatfix,showpacs]{revtex4-1}

\usepackage{amsthm}
\usepackage{amsmath}
\usepackage{graphicx}
\usepackage{amssymb}
\usepackage{bm}
\usepackage{latexsym}
\usepackage{babel}
\usepackage[bookmarks=false]{hyperref}

\begin{document}

\author{F. Mulansky}
\affiliation{Department of Physics and Astronomy, McMaster University, 1280 Main St.\ W., Hamilton, ON, L8S 4M1, Canada}
\author{J. Mumford}
\affiliation{Department of Physics and Astronomy, McMaster University, 1280 Main St.\ W., Hamilton, ON, L8S 4M1, Canada}
\author{D. H. J. O'Dell}
\affiliation{Department of Physics and Astronomy, McMaster University, 1280 Main St.\ W., Hamilton, ON, L8S 4M1, Canada}

\title{Impurity in a Bose-Einstein condensate in a double well}

\begin{abstract}
We compare and contrast the mean-field and many-body properties of a Bose-Einstein condensate trapped in a double well potential with a single impurity atom. The mean-field solutions display a rich structure of bifurcations as parameters such as the boson-impurity interaction strength and the tilt between the two wells are varied. In particular, we study a pitchfork bifurcation in the lowest mean-field stationary solution which occurs when the boson-impurity interaction exceeds a critical magnitude. This bifurcation, which is present for both repulsive and attractive boson-impurity interactions, corresponds to the spontaneous formation of an imbalance in the number of particles between the two wells. If the boson-impurity interaction is large, the bifurcation is associated with the onset of a Schr\"{o}dinger cat state in the many-body ground state. We calculate the coherence and number fluctuations between the two wells, and also the entanglement entropy between the bosons and the impurity. We find that the coherence can be greatly enhanced at the bifurcation.
\end{abstract}

\pacs{03.75.Lm, 05.45.Mt, 03.75.Gg, 03.65.Ta}

\maketitle

\section{Introduction}
\label{sec:introduction}

In recent years a large number of experiments have been performed upon atomic Bose-Einstein condensates (BECs) trapped in double well potentials, see e.g., \cite{andrews97,shin04,wang05,albiez05,schumm05,jo07,hofferberth07,levy07,esteve08,shomroni09,maussang10,baumgartner10,leblanc10,chen11}.
Each atom can be in a superposition of being in both wells, allowing for studies of macroscopic quantum coherence \cite{javanainen86,jack96,andrews97,shin04,wang05,schumm05,burkov07,stimming11} analogous to that found in a Josephson junction \cite{smerzi97,leggett01,P&S,albiez05,levy07,gati07,leblanc10}. However, in contrast to a traditional solid state Josephson junction, the microscopic hamiltonian of an atomic gas is highly controllable and is very well understood. The double well system is also one of the simplest ways to go beyond the mean-field Gross-Pitaevskii paradigm of gaseous BECs because the inter-well tunnelling introduces a very low energy scale which can easily be surpassed by the interaction energy. This tends to emphasize the discrete (second-quantized) aspects of the quantum state, which are conjugate to the phase properties, and can be seen in effects such as number squeezing \cite{jo07,esteve08}. Double well systems are, of course, also naturally disposed to being used for matter-wave interferometery \cite{shin04,wang05,schumm05,baumgartner10}, with the eventual aim of making precision measurements.

In this paper we consider $N$ identical bosons trapped in a double well potential, and model this system via the two-site Bose-Hubbard model. To this we add a single impurity atom. The impurity can be an atom of a different species or internal state, but we assume that it is also trapped in the double well potential (but need not have the same tunneling rate between the wells as the bosons). Reference \cite{cirone09} considers the related, but different, case of an impurity atom trapped in a double well potential immersed in a uniform BEC as an example of the spin-boson model.  In our model the bosons are limited to just two states (two-mode approximation), whereas in a uniform BEC there is a continuum of bosonic states. Another study which is related to ours considers an atomic quantum dot acting as a coherent single atom shuttle between two BECs \cite{bausmerth07}.

One motivation for studying an impurity interacting with a BEC in a double well is that it can be used as a simple, yet concrete, model for studying the measurement problem in quantum mechanics \cite{spehner}: the impurity is a two-state quantum system interacting with a macroscopic measurement device represented by the bosons. The measurement device can be tuned between being quantum ($N$ small) or classical ($N$ large). Furthermore, the use of a Feshbach resonance would allow the interaction between the quantum system and the measurement device to be tuned between being weak and strong. In this context we note that the interesting question of the classical (mean-field) limit of bosons in double well potential has been addressed in a number of papers \cite{wu06, krahn09,chuchem10}  with the general conclusion that this occurs when $N \rightarrow \infty$, although with the caveat that quantum effects remain important close to the separatrix \cite{krahn09,chuchem10}, which is the boundary in phase space outside of which self-trapping occurs.

A theoretical study of the same system as considered here has been performed by Rinck and Bruder \cite{rinck11}. Their paper, which focuses on the case of small atom number, predicts that when the boson-impurity interaction is strong a tunnelling resonance occurs that involves the simultaneous tunnelling of many bosons together with the impurity. In the presence of a tilt asymmetry between the two wells this resonance corresponds to the expulsion of the impurity into the higher lying well. In this paper we extend their study to larger particle numbers and consider both the full many-body theory and the Gross-Pitaevskii mean-field theory (which is expected to be valid in the large $N$ regime). We find that the Rinck-Bruder tunnelling resonance is associated with   
a bifurcation in the mean-field theory solutions.

Although it might seem rather fanciful to study the problem of a single impurity in a BEC in a double well, we note that a recent experiment \cite{will11} has realized an optical lattice with many bosons and one impurity per site, which is quite close to the situation we are considering here.

\section{The many-body hamiltonian}
\label{sec:manybody}

We begin by writing down the many-body hamiltonian for $N$ identical bosons interacting with a single impurity. Both the bosons and the impurity are trapped in a double well potential (for simplicity we shall assume in later sections that the potential is the same for both). The impurity is a particle distinguishable from the bosons---it may be either a boson or a fermion, but its statistics do not matter. It can even be the same species of atom as the bosons, but in a different hyperfine state provided there is no interconversion between the hyperfine states. We then make the two-mode approximation (i.e.\ the single band approximation for the two-site Bose-Hubbard model) for the bosons and likewise for the impurity, so that there are four modes and $2 \times (N+1)$ many-body states in total.

The total hamiltonian for the BEC plus impurity system is
\begin{equation}
	\widehat{H} = \widehat{H}_A + \widehat{H}_B + \widehat{H}_{BB} + \widehat{H}_{AB}
	\label{eq:totalhamiltonian}
\end{equation}
where $\widehat{H}_A$ is the single-particle hamiltonian for the impurity, $\widehat{H}_B$ is the single-particle hamiltonian for the bosons, $\widehat{H}_{BB}$ is the boson-boson interaction hamiltonian, and $\widehat{H}_{AB}$ is boson-impurity interaction hamiltonian. In terms of the field operators $\hat{\Phi}(x)$ for the bosons and  $\hat{\Psi}(x)$ for the impurity, we have  
\begin{eqnarray}
	\widehat{H}_A &=& \int d^3x \hat{\Psi}^{\dag}(x) \left[-\frac{\hbar^2}{2m_A} \nabla^2+V_A(x) \right] \hat{\Psi}(x) \label{eq:impurityhamiltonian}  \\
	\widehat{H}_B &=& \int d^3x \hat{\Phi}^{\dag}(x) \left[-\frac{\hbar^2}{2m_B} \nabla^2+V_B(x) \right] \hat{\Phi}(x) \label{eq:bosonhamiltonian}\\
	\widehat{H}_{BB} &=& \frac{g_{BB}}{2}\int d^3x \hat{\Phi}^{\dag}(x)\hat{\Phi}^{\dag}(x)\hat{\Phi}(x)\hat{\Phi}(x) \label{eq:bosonbosonhamiltonian} \\ & &~~~~~~ \text{(boson-boson interaction)} \nonumber \\
	\widehat{H}_{AB} &=& g_{AB}\int d^3x  \hat{\Psi}^{\dag} (x)  \hat{\Phi}^{\dag}(x)\hat{\Phi}(x)\hat{\Psi}(x) \label{eq:bosonimpurityhamiltonian} \\ && ~~~~ \text{(boson-impurity interaction)} \nonumber
\end{eqnarray}
where $g_{BB} \equiv 4 \pi \hbar^2 a_{BB} /m_{B}$ characterizes the boson-boson interactions in terms of the inter-boson $s$-wave scattering length $a_{BB}$ and the boson mass $m_{B}$, and $g_{AB} \equiv  2 \pi \hbar^2 a_{AB} /m_{r}$ characterizes the boson-impurity interactions in terms of the boson-impurity $s$-wave scattering length $a_{AB}$ and the reduced mass $m_{r} = m_{A}m_{B}/(m_{A}+m_{B})$, with $m_{A}$ being the mass of the impurity atom. The field operators in the above expressions obey the commutation relations $[\hat{\Phi}(x),\hat{\Phi}^{\dag}(x')]=\delta(x-x')$, $[\hat{\Psi}(x),\hat{\Psi}^{\dag}(x')]=\delta(x-x')$,  $[\hat{\Phi}(x),\hat{\Psi}^{\dag}(x')]=0$, and $[\hat{\Phi}(x),\hat{\Psi}(x')]=0$. 

In this paper we shall work in the two-mode approximation. For each species the ground mode is symmetric and the excited mode is antisymmetric: the energy splitting between them is determined by tunneling and is therefore exponentially smaller than the energy separation to higher modes, allowing us to isolate just the lowest two modes. Even and odd combinations of the symmetric and antisymmetric modes give two new orthogonal modes which are localized, respectively, on the left and right hand sides of the double well: $\phi_{i}(x)$ for the bosons and $\psi_{i}(x)$ for the impurity, with $\int \phi_{i}^{*}\phi_{j} d^{3}x= \int \psi_{i}^{*}\psi_{j} d^{3}x = \delta_{ij}$, where $i$ and $j$ can be either $L$ or $R$.
Expanding the field operators in terms of the localized modes we have
\begin{eqnarray}
\hat{\Psi}(x) &=& \hat{a}_L \psi_L(x) + \hat{a}_R \psi_R(x) \label{eq:twomodea} \\
\hat{\Phi}(x)&=& \hat{b}_L \phi_L(x) + \hat{b}_R \phi_R(x) \, , \label{eq:twomodeb}
\end{eqnarray}
where the mode operators for the bosons obey the commutation relations $[\hat{b}_{i},\hat{b}_{j}^{\dag}]=\delta_{ij}$, and $[\hat{b}_{i},\hat{b}_{j}]=0$. A similar set of commutation relations hold for the impurity operators, and the boson and impurity operators commute with each other.  
Substituting these expansions into the hamiltonian we obtain the two-mode hamiltonian, which can be written, up to constant terms, as \cite{rinck11}
\begin{eqnarray}
\widehat{H} & = &  \frac{U}{4} {\Delta \hat N}^2  - J \hat{B} - J^a \hat{A}  \nonumber \\ & &  + \frac{W}{2} \Delta\hat N \Delta\hat M + \frac{\Delta \epsilon}{2} \Delta\hat{N} + \frac{\Delta \epsilon^a}{2} \Delta\hat{M}
\label{eq:twomodehamiltonian}
\end{eqnarray}
where we have defined: $\Delta \hat{M} \equiv \hat{a}_R^{\dag}\hat{a}_R-\hat{a}_L^{\dag}\hat{a}_L$ as the number difference operator between the two wells for the impurity (its eigenvalues are $\pm 1$); $\Delta \hat{N}\equiv \hat{b}_R^{\dag}\hat{b}_R-\hat{b}_L^{\dag}\hat{b}_L$  ditto for the bosons (its eigenvalues range from $-N$ to $N$ in steps of two); and $\hat{A} \equiv \hat{a}_L^{\dag}\hat{a}_R + \hat{a}_R^{\dag}\hat{a}_L$ and $\hat{B} \equiv \hat{b}_L^{\dag}\hat{b}_R + \hat{b}_R^{\dag}\hat{b}_L$ as the hopping operators for the impurity and bosons, respectively. The parameters that appear in the two-mode hamiltonian are defined as 
\begin{eqnarray}
	U_{L,R} & \equiv & g_{BB} \int d^3x |\phi_{L,R}|^4 \label{eq:Udefinition} \\
	W_{L,R} & \equiv & g_{AB} \int d^3x |\phi_{L,R}|^2|\psi_{L,R}|^2  \label{eq:Wdefinition} \\
	J & \equiv & -\int d^3x \phi_{L,R}^*\left[-\frac{\hbar^2}{2m_B} \nabla^2+V_B(x)\right]\phi_{R,L}  \label{eq:Jdefinition} \\
	\Delta \epsilon & \equiv & \epsilon_{R} - \epsilon_{L} \, .  \label{eq:Deltaepsilondefinition}
\end{eqnarray}
$U_{L/R}$ is the intra-well interaction energy for the bosons, and $W_{L/R}$ is the intra-well interaction energy between the bosons and the impurity. For the rest of this paper we shall assume that $U_{L}=U_{R}=U$ and $W_{L}=W_{R}=W$.   $J$ and $\Delta \epsilon$ are, respectively, the bosonic hopping energy and difference in zero-point single-particle energies between the two wells.  $J^{a}$ and $\Delta \epsilon^{a}$ are the equivalent quantities for the impurity. The single-particle energy $\epsilon_{L,R}$ for the bosons is defined as
\begin{equation}
\epsilon_{L,R}  \equiv  \int d^3x \phi_{L,R}^*\left[-\frac{\hbar^2}{2m_B} \nabla^2+V_B(x)\right]\phi_{L,R} \, .
\end{equation}
The quantity $\Delta \epsilon$ can be regarded as an imbalance or tilt between the two wells arising, e.g., due to gravity if one well is  lower than the other.

Note that the expression (\ref{eq:twomodehamiltonian}) for the two-mode hamiltonian neglects small cross terms such as those that depend on integrals like $\int d^{3}x \phi_{L}^{*}\phi_{L} \psi_{L}^{*} \psi_{R}$. Furthermore, we shall also assume that the trapping provided by the wells is tight enough that the single-particle energies $\epsilon_{L,R}$ dominate the interaction energies. This allows us to neglect changes in the mode wave functions, and hence changes in the parameters (\ref{eq:Udefinition})--(\ref{eq:Deltaepsilondefinition}), as the particle number in each well varies.

Let us find the matrix elements of $\widehat{H}$ in the Fock basis $\vert M_{R},N_{R} \rangle$, where $M_{R}$ is the eigenvalue of the impurity number operator for the right well $\hat{M}_{R} \equiv \hat{a}_{R}^{\dag}\hat{a}_{R}$, and $N_{R}$ is the equivalent eigenvalue for the bosons. Due to number conservation, we only need to specify the number of particles in the right hand well. We find that the general matrix element is
\begin{eqnarray}
	\langle M_R^\prime N_R^\prime \vert &  \widehat{H} & \vert M_R N_R\rangle = \nonumber \\  \Big( \frac{\Delta \epsilon}{2} \Delta N &+& \frac{U}{4} {\Delta N}^2 + \frac{\Delta \epsilon^a}{2} \Delta M + \frac{W}{2} \Delta N \Delta M \Big) \delta_{N_R^\prime N_R} \delta_{M_R^\prime M_R} \nonumber \\
	&& -J \sqrt{N_R(N-N_R+1)} \delta_{N_R^\prime N_R-1} \delta_{M_R^\prime M_R} \nonumber \\
	&& -J \sqrt{(N-N_R)(N_R+1)} \delta_{N_R^\prime N_R+1} \delta_{M_R^\prime M_R} \nonumber \\
	&& -J^a \sqrt{M_R(M-M_R+1)} \delta_{N_R^\prime N_R} \delta_{M_R^\prime M_R-1} \nonumber \\
	&& -J^a \sqrt{(M-M_R)(M_R+1)} \delta_{N_R^\prime N_R} \delta_{M_R^\prime M_R+1}  \label{eq:matrixelement}
\end{eqnarray}
where $M=1$ is the number of impurities, $\Delta N = N_{R}-N_{L}$ and $\Delta M= M_{R}-M_{L}$.

\begin{figure}
	\includegraphics[width=0.9\columnwidth]{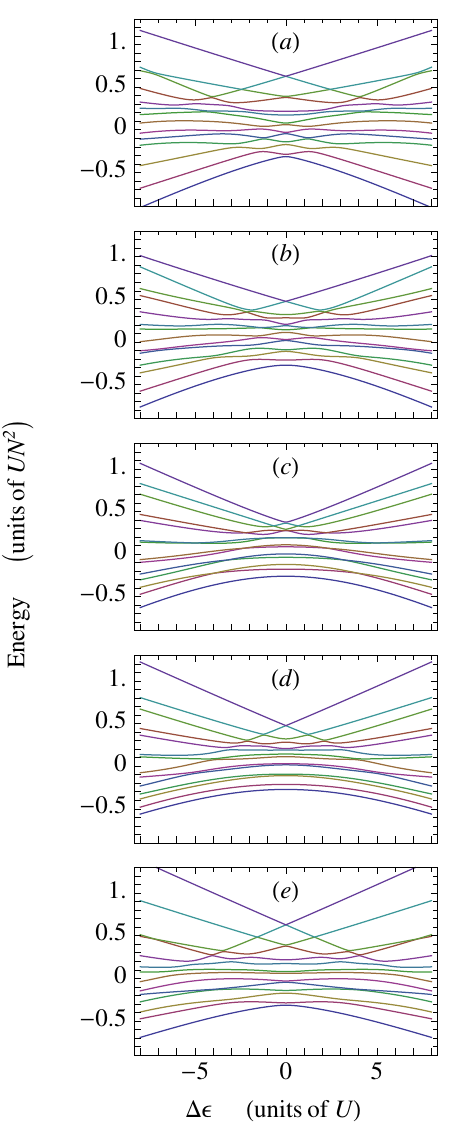}
	\caption{(Color online) Eigenvalues of the many-body hamiltonian as a function of the tilt $\Delta \epsilon$ $(=\Delta \epsilon^{a})$ for $N=6$ bosons and a single impurity. Each panel has a different value of the boson-impurity coupling:  (a) $W/U=-4$, (b) $W/U=-2$, (c) $W/U=0$, (d) $W/U=2$, (e) $W/U=4$. The values of the other parameters are $J/U=J^{a}/U=1.5$.}
	\label{fig:eigenvalues}
\end{figure}

Diagonalizing the matrix specified by Eq.\ (\ref{eq:matrixelement}) gives the eigenvalues and eigenvectors of the many-body two-mode hamiltonian. In Figure \ref{fig:eigenvalues} we plot the eigenvalues as a function of the tilt $\Delta \epsilon$ for the case of six bosons for both positive and negative values of the boson-impurity interaction $W$. The avoided crossings correspond to places where a particle hops from one well to another as the tilt is changed. Similar pictures have previously been made for the BEC-impurity system by Rinck and Bruder \cite{rinck11}, who noted that when the boson-impurity interaction $W$ is repulsive enough the impurity can be forced to tunnel uphill against the gradient set by $\Delta \epsilon^{a}$, i.e.\ the impurity can be expelled from the BEC into the higher lying well.

The problem of a BEC in a double well potential can be mapped onto a pendulum model, or, equivalently, a single particle in a periodic potential. As explained in \cite{krahn09}, in this latter picture the tilt $\Delta \epsilon$ plays the role of the quasimomentum of the particle, and the eigenvalue structure seen in Fig.\ \ref{fig:eigenvalues} as a function of tilt can then be viewed as a band structure plotted as a function of quasimomentum. The energy level structure shown in Fig.\ \ref{fig:eigenvalues} does not have the usual periodic form we expect of a band structure, but it can be made periodic by applying a simple transformation \cite{krahn09}. 

\section{Mean-field approximation}
\label{sec:mfa}

Let us now use the mean-field approximation to calculate the equations of motion for the number and phase differences of the condensate and impurity wave functions between the two wells.  The stationary solutions of these equations provide natural quantities to compare with the eigenvalue structure of the second-quantized theory because both quantities are stationary during time evolution.
The relationship between the second quantized theory presented in Section \ref{sec:manybody},  and the mean-field theory presented in this section, is analogous to that between quantum and classical mechanics for a single particle: the $N \rightarrow \infty$ limit of the many-body theory gives the mean-field theory, and is equivalent to the $\hbar \rightarrow 0$ limit of single particle quantum theory \cite{wu06}.  In fact, the precise relationship is $NJ/U=(S/\hbar)^2$ where $S$ is the classical action \cite{krahn09}.

In the presence of Bose-Einstein condensation the bosonic field operator can be written $\hat{\Phi}(x)=\Phi_{0}(x)+\delta \hat{\Phi}(x)$, where $\Phi_{0}(x)$ is the so-called condensate wave function. A description of the bosons solely in terms of $\Phi_{0}$ corresponds to a mean-field approximation. Making this  replacement in 
the hamiltonian (\ref{eq:totalhamiltonian}) gives the mean-field energy functional $H$ (because there is only one impurity atom, we can also replace its field operator by its wave function $\hat{\Psi} \rightarrow \Psi$ without involving any approximation).  
The equations of motion for the impurity and condensate wave functions can be found by taking functional derivatives of $H$, i.e.\ $i \hbar \partial \Psi/\partial t = \delta H/ \delta \Psi^{*}$ and $i \hbar \partial \Phi_{0}/\partial t = \delta H/ \delta \Phi_{0}^{*}$. The result is a Schr\"{o}dinger equation for the impurity
\begin{equation}
i\hbar \frac{\partial \Psi}{\partial t}  =  \left[-\frac{\hbar^2}{2m_A} \nabla^2+V_A(x) + g_{AB}
	\left\vert \Phi_{0} \right\vert^{2} \right] \Psi 
	\label{eq:schrod}
\end{equation}
coupled to a Gross-Pitaevskii equation for the bosons
\begin{equation}
	i\hbar \frac{\partial \Phi_{0}}{\partial t}  =  \left[-\frac{\hbar^2}{2m_B} \nabla^2+V_B(x) +g_B \left\vert\Phi_{0}\right\vert^2 + g_{AB} \left\vert \Psi \right\vert^2 \right] \Phi_{0} \, .
	\label{eq:GPE} 
\end{equation}

In order to make the two-mode approximation in the mean-field case, we replace the mode operators $\hat{a}_{L/R}$ and $\hat{b}_{L/R}$ that appear in Eqns.\ (\ref{eq:twomodea}) and (\ref{eq:twomodeb}) by the complex numbers $a_{L/R}$ and $b_{L/R}$
\begin{eqnarray}
a_{L/R} &=& \sqrt{M_{L/R}}e^{i\alpha_{L/R}(t)} \\
	b_{L/R}&=& \sqrt{N_{L/R}}e^{i\beta_{L/R}(t)}  \, .
\end{eqnarray}
Substituting these forms into the Schr\"{o}dinger and Gross-Pitaevskii equations we obtain the equations of motion 
\begin{eqnarray}
 \dot{\alpha} &= & \frac{\Delta\epsilon^a}{\hbar} + 2\frac{W}{\hbar} Z + \frac{4J^a}{\hbar}  \frac{Y \cos{\alpha}}{\sqrt{1-4Y^{2}}} \label{eq:josephson1} \\
	 \dot{Y} &= & -\frac{J^a}{\hbar} \sqrt{1-4 Y^{2}} \sin{\alpha} \label{eq:josephson2} \\
	\dot{\beta} &= & \frac{\Delta\epsilon}{\hbar} + 2\frac{U}{\hbar} Z + 2\frac{W}{\hbar} Y + \frac{4J}{\hbar} \frac{ Z \cos{\beta} }{\sqrt{N^2-4Z^{2}}} \label{eq:josephson3} \\
	  \dot{Z} &= & - \frac{J}{\hbar} \sqrt{N^2-4 Z^{2}} \sin{\beta}  \label{eq:josephson4} 
	 \end{eqnarray}
where we have defined the variables $\alpha \equiv \alpha_{L}-\alpha_{R}$, $Y \equiv \Delta M/2$, $\beta \equiv \beta_{L}-\beta_{R}$, and $Z \equiv \Delta N/2$. Recognizing that the canonically conjugate pairs of variables are $\{\alpha,\hbar Y \}$ and $\{\beta, \hbar Z\}$, the equations of motion can be expressed in the form of Hamilton's equations
\begin{eqnarray}
\dot{\alpha} = \frac{1}{\hbar} \frac{\partial H}{\partial Y}  \quad ; \quad \dot{Y} =  - \frac{1}{\hbar} \frac{\partial H}{\partial \alpha} \\
\dot{\beta} = \frac{1}{\hbar} \frac{\partial H}{\partial Z}  \quad ; \quad  \dot{Z} =  - \frac{1}{\hbar}\frac{\partial H}{\partial \beta}
\end{eqnarray}
where $H$ in terms of the new variables is
\begin{eqnarray}
	H &=& UZ^{2} - J  \sqrt{N^2-4Z^{2}} \cos{\beta}  - J^a  \sqrt{1- 4 Y^2} \cos{\alpha} \nonumber  \\
	&&  + 2 W Y Z + \Delta\epsilon \, Z + \Delta\epsilon^a \, Y \, . \label{eq:mfahamiltonian}
\end{eqnarray}
This mean-field hamiltonian function can be compared term by term with the second-quantized version given in Eq.\ (\ref{eq:twomodehamiltonian}).

In order to gain some intuition, let us first consider the case where $W=0$, so that the impurity and the BEC are not coupled to each other. The impurity is an elementary two state system, analogous to, e.g., a spin or a two-level atom. The equations of motion for the impurity can be solved exactly to give
\begin{eqnarray}
Y(t) & = & Y_{0} \sin \left(\omega_{\mathrm{imp}} t + \phi_{a} \right) \\
\alpha(t) & = & \arcsin \left[\frac{ -2 Y_{0} \cos (\omega_{\mathrm{imp}}t + \phi_{a})}{\sqrt{1-4Y_{0}^2 \sin^{2} (\omega_{\mathrm{imp}}t+\phi_{a})}}\right]
\end{eqnarray}
where the constant $\phi_{a}$ gives the initial phase of the motion, the constant $Y_{0}$ sets the magnitude of the oscillations of the amplitude $Y(t)$, and must lie in the range $-1/2 \leq Y_{0} \leq 1/2$, and the angular frequency $\omega_{\mathrm{imp}}=2 J^{a}/\hbar$ of the oscillation is given by the bare hopping frequency. For simplicity we have assumed that the tilt $\Delta \epsilon^{a}/\hbar=0$. 

When $W \neq 0$, the coupled motion of the impurity and the BEC can be complicated (indeed, it is expected to be chaotic---see later). A simplified situation arises if the BEC is static, i.e.\ if $J=0$ so that its tunneling is switched off and $Z$ is locked at a particular value.  We may then ask, how does the presence of the BEC affect the tunneling frequency of the impurity? The answer is, not very much. As can be seen by inspection of Eqns. (\ref{eq:josephson1})--(\ref{eq:josephson4}), the effect of a static BEC upon the impurity is exactly the same as the tilt term containing $\Delta \epsilon^{a}$, which does not affect the frequency of motion. In fact, if $Z=0$ the impurity is completely unaffected by the BEC. Thus, the impurity does not acquire an `effective mass' through its interaction with the BEC. This result probably only holds in the Bose-Hubbard limit considered in this paper. When the mode wave functions are allowed to be modified by interactions it is likely that the impurity will acquire an `effective mass' which will affect $\omega_{\mathrm{imp}}$.

Let us now turn to the BEC component. In the absence of the impurity, the equations of motion  (\ref{eq:josephson3})  and (\ref{eq:josephson4})  for the BEC correspond to the celebrated Josephson equations \cite{josephson62,P&S} known from the theory of superconductivity.  Providing the atom number imbalance between the two wells is much smaller than the total number $N$ (low energy regime), and the inequality  $NU \gg J$ is obeyed \cite{P&S},  we can replace $\sqrt{N^2-4Z^2}$ in the mean-field hamiltonian by $N$  to leave
\begin{equation} 
H_{\mathrm{BEC}} \approx UZ^{2} - J  N \cos{\beta} \, . \label{eq:mfahamiltonianBEC}
\end{equation}
For small angles, this yields the equations of motion
\begin{eqnarray}
\dot{\beta} & = & \frac{2U}{\hbar} Z \\
\dot{Z} & = & -\frac{JN}{\hbar} \beta
\end{eqnarray}
whose solutions are
\begin{eqnarray}
Z(t)  & = & Z_{0} \sin \left(\omega_{\mathrm{plas}} t + \phi_{z} \right) \\
\beta(t) & = & \beta_{0} \sin \left(\omega_{\mathrm{plas}} t + \phi_{\beta} \right) \, .
\end{eqnarray}
The constants $\beta_{0}$, $\phi_{\beta}$, $Z_{0}$ and $\phi_{z}$ are set by the boundary conditions, and the amplitude of the number difference oscillations must lie in the range $N/2 \leq Z_{0} \le N/2$. The frequency of the oscillations predicted by the above equations of motion is 
\begin{equation}
\omega_{\mathrm{plas}} =\frac{ \sqrt{2 J U N}}{\hbar} \, , \label{eq:plasmafrequency}
\end{equation}
and is known as the plasma frequency. We see that, due to the interactions between the bosons, the frequency of their oscillation depends on their total number $N$ and hence can, counter-intuitively, be much faster than that of the impurity (assuming all the parameters for the impurity are chosen to be equal to their counterparts for the bosons). 

The hamiltonian (\ref{eq:mfahamiltonianBEC}) for the BEC has the same form as that of a pendulum
\begin{equation}
H_{\mathrm{pend}}=\frac{p^2}{2ml^2}-mgl \cos \beta
\end{equation}
where $m$ is the mass of the bob, $l$ the length of the (massless) rod, and $g$ is the acceleration due to gravity. For the case of the pendulum, $\beta$ becomes  the angular displacement from the downward vertical, and it is conjugate to the angular momentum $p$. The analogy between the pendulum and the BEC is completed by relating $p$ to the population imbalance via $p = \hbar Z$. 

From the position of the factor $N$ in the `potential energy' term in Eq.\ (\ref{eq:mfahamiltonianBEC}), it is tempting to associate it with the length $l$ of the pendulum, so that $l \propto N$ \cite{smerzi97,albiez05}. However, this is wrong because it implies that the frequency $\omega_{\mathrm{pend}}=\sqrt{g/l}$ of the pendulum reduces with increasing $N$, in contradiction to what $\omega_{\mathrm{plas}}$ predicts. Because $N$ is dimensionless, we can instead let $N=L/l$, where $L$ is a constant with units of length, e.g. the length corresponding to N=1. Then the length of the effective pendulum is inversely related to the number of atoms and $\omega_{\mathrm{pend}}=\sqrt{N g/L}$, as required. A similar approach can be applied to the more general case of Eq.\ (\ref{eq:mfahamiltonian}), where the terms $\sqrt{N^2-4 Z^2}$ imply a pendulum whose length changes during the motion \cite{smerzi97}. The length of this pendulum increases with $Z$.

The above considerations lead us to a picture for the combined BEC-impurity system as two coupled pendula with natural frequencies $\omega_{1}=\sqrt{g/l}$ and $\omega_{2}=\sqrt{g/L}$, where $l \propto 1/N$. When $N \gg 1$ the length of the BEC pendulum is much shorter than that of the impurity pendulum so that $l \ll L$. The coupling $2 W Y Z$ between the two pendula depends upon the product of the two angular momenta. This implies that we cannot think of two pendula coupled by a spring, because that would lead to a coupling term that depends on the difference in the angles $\alpha - \beta$. An alternative model which does have a coupling of the correct form is the double pendulum, where one pendulum is suspended from the other. 
The hamiltonian for the double pendulum is derived in the Appendix and can be written (when $m_{1} \gg m_{2}$)
\begin{eqnarray}
H_{\mathrm{dp}} & = & \frac{1}{2} \left\{ \frac{p_{1}^{2}}{m_{1} l_{1}^2 }+\frac{p_{2}^{2}}{m_{2} l_{2}^{2}}-\frac{2}{m_{1}}\frac{p_{1}}{l_{1}}\frac{p_{2}}{l_{2}} \cos(\theta_{1}-\theta_{2}) \right\} \nonumber \\
 & & - m_{1}gl_{1} \cos \theta_{1}-m_{2}g l_{2} \cos \theta_{2} \, .
\end{eqnarray}
where $p_{1}$ and $p_{2}$ are the angular momenta conjugate to the angles $\theta_{1}$ and $\theta_{2}$.  For small deviations from the stationary solutions the $\cos (\theta_{1}-\theta_{2})$ term is approximately constant. For example, when both pendula are pointing down $\theta_{1}=\theta_{2}=0$, so that $\cos (\theta_{1}-\theta_{2})=1$. When one pendulum is pointing down and the other is pointing up $\cos (\theta_{1}-\theta_{2})=-1$. In these situations the double pendulum model serves as a qualitative model which can help guide our intuition  for the BEC-impurity system. Indeed, a well known feature of the double pendulum is that it can display chaotic motion, and we expect this to also be true of the BEC-impurity system in the mean-field regime.
However, one shortcoming of the double-pendulum model is that one cannot choose the strength of the coupling independently of the properties of the individual pendula.

\section{Static solutions to the Mean-Field Equations: swallowtail loops and pitchfork bifurcations}
\label{sec:staticsolutions}

\begin{figure}
\includegraphics[width=0.9\columnwidth]{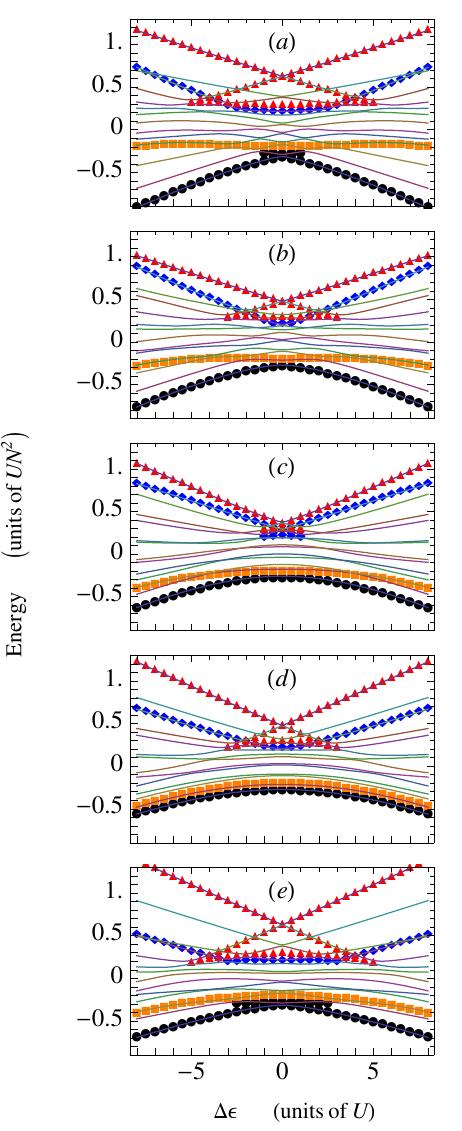}
\caption{(Color online) Energies of the static solutions to the mean-field equations (\ref{eq:josephson1})--(\ref{eq:josephson4}) as a function of the tilt $\Delta \epsilon$ ($=\Delta \epsilon^{a}$).  The various solutions are characterized by their phase differences: $\alpha=\beta=0$ (black circles); $\alpha=\pi$, $\beta=0$ (orange squares); $\alpha=0$, $\beta=\pi$ (blue diamonds); and  $\alpha=\beta=\pi$ (red triangles). Each panel has a different value of the boson-impurity coupling:  (a) $W/U=-4$, (b) $W/U=-2$, (c) $W/U=0$, (d) $W/U=2$, (e) $W/U=4$. All panels have $J/U=J^{a}/U=1.5$. We have also included the many-body eigenvalues (thin lines) for  $N=6$ bosons and a single impurity.}
\label{fig:loopwithtilt}
\end{figure}

\begin{figure}
		\includegraphics[width=0.9\columnwidth]{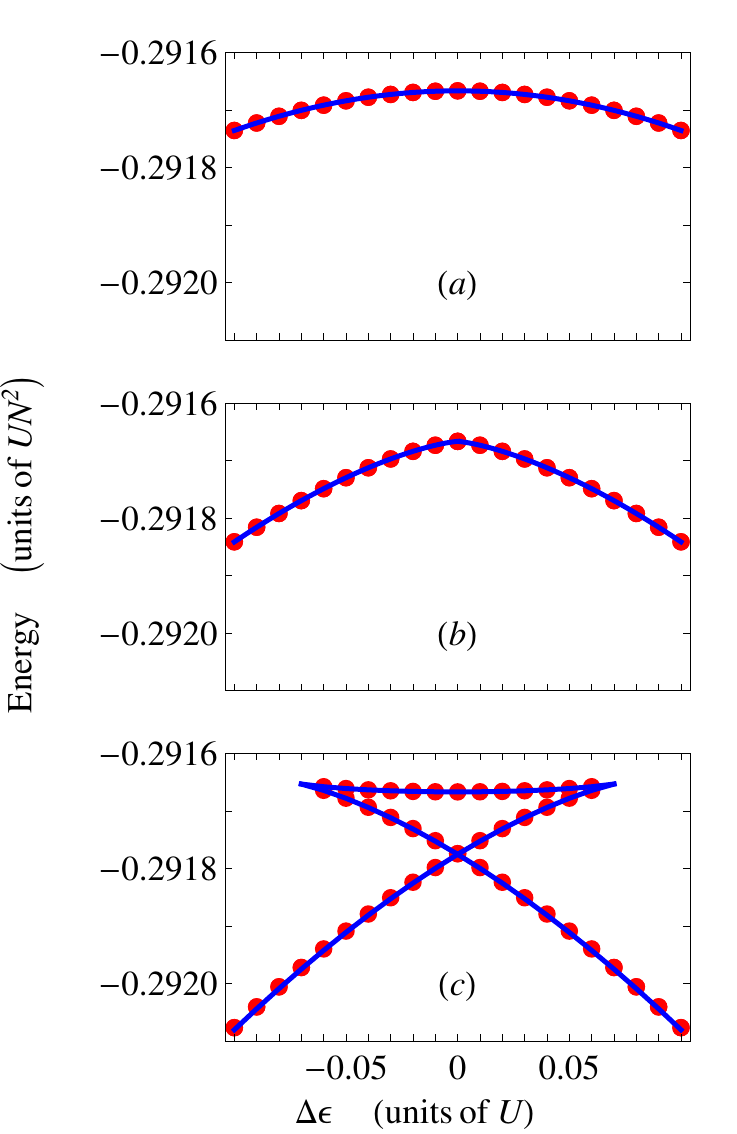}
	\caption{(Color online) A zoom-in of Fig.\ \ref{fig:loopwithtilt} showing the birth of the loop in the lowest band as the boson-impurity interaction strength $W$ is varied: (a) $W/U=2.0$; (b) $W/U=2.122$; (c) W/U=2.2. The solid lines are given by solving $F(\theta)=0$, as given in Eq.~(\ref{eq:thetafunction}), and the dots are given by solving the 3rd order Taylor expansion $F_{3}(\theta)$, as given in Eq.~(\ref{eq:taylorexpansion}). The other parameters are set at $N=6$, $J=J^{a}=1.5 \, U$.}
\label{fig:groundloop}
\end{figure}

\begin{figure}
		\includegraphics[width=0.9\columnwidth]{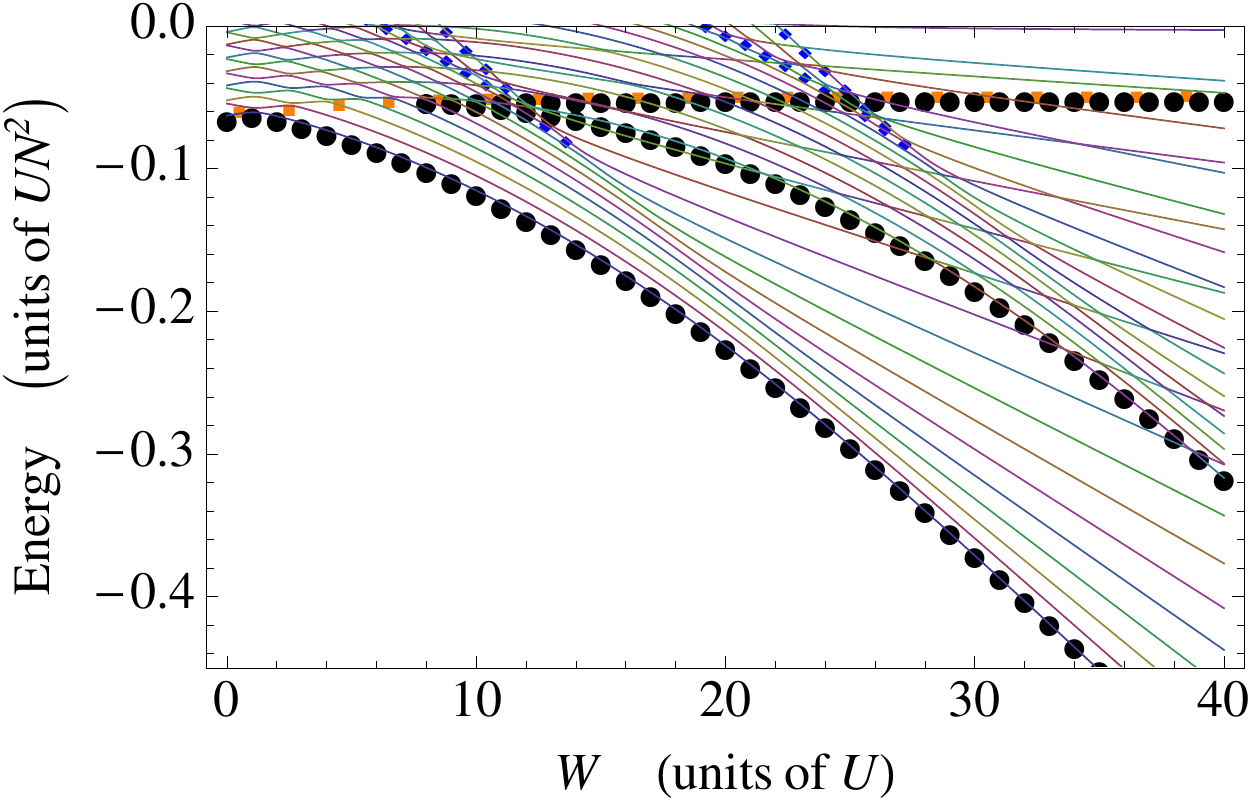}
	\caption{(Color online) Dependence of the many-body and mean-field energies upon the boson-impurity interaction $W$ for a finite value of the tilt. The solid curves are many-body eigenvalues, and the dotted curves are the static solutions to the mean-field equations: $\alpha=\beta=0$ (black circles); $\alpha=\pi$, $\beta=0$ (orange squares); $\alpha=0$, $\beta=\pi$ (blue diamonds). This figure only shows the low energies and so the $\alpha=\beta=\pi$ solution does not appear. The other parameters are $N=30$, $J=J^{a}=1.5 \, U$, $\Delta \epsilon = \Delta \epsilon^{a}=7 \, U$.}
	\label{fig:pitchforkenergy}
\end{figure}

\begin{figure}
		\includegraphics[width=0.9\columnwidth]{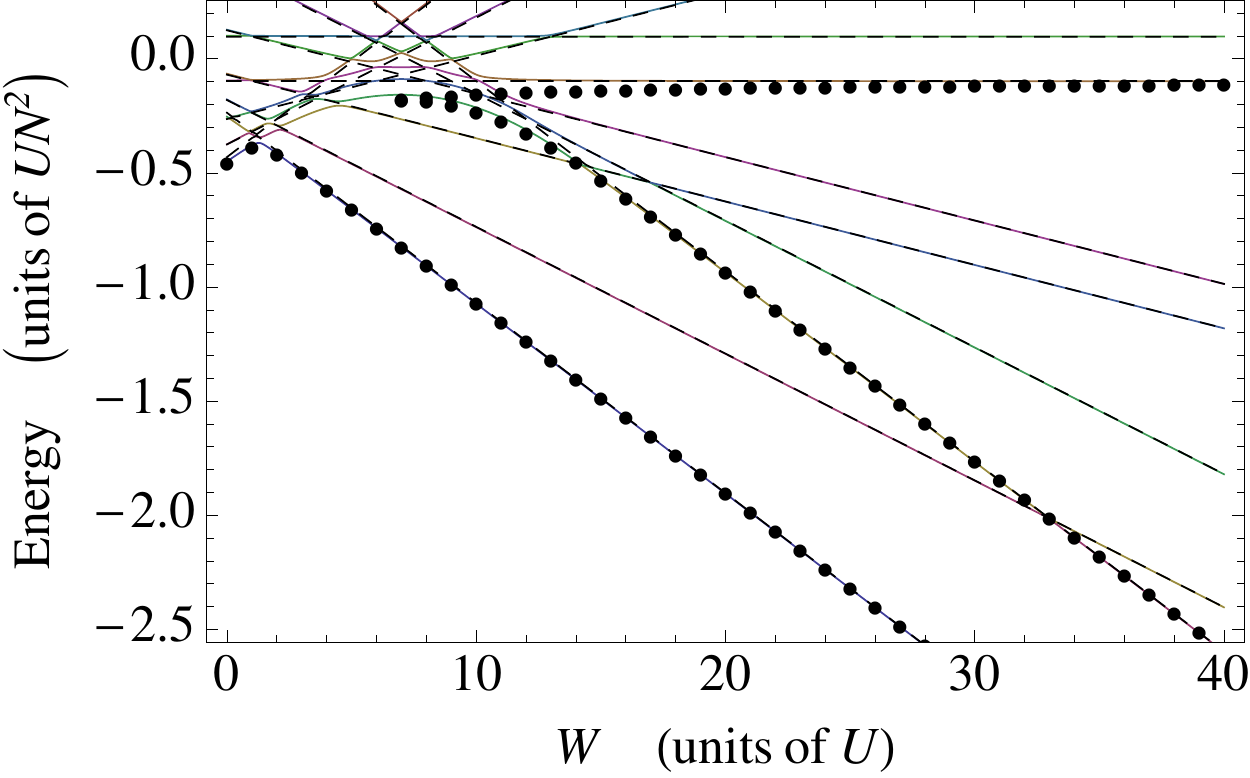}
	\caption{(Color online) Dependence of the many-body and mean-field energies upon the boson-impurity interaction $W$ for a finite value of the tilt $\Delta \epsilon = \Delta \epsilon^{a}=7 \, U$. This picture differs from Fig.\ \ref{fig:pitchforkenergy} in that we have far fewer atoms ($N=6$), and we have decreased the ratio of $J$ to $J=J^{a}=0.5 \, U$. We have also plotted the energies of the relevant Fock states (dashed lines). The solid curves, which mostly lie on top of the Fock state energies, are the eigenvalues of the many-body hamiltonian.  The black circles give the $\alpha=\beta=0$ static solutions to the mean-field equations (for simplicity we have not shown the other branches). }
	\label{fig:pitchsmallN}
\end{figure}

In this section we compute the static solutions to the mean-field equations of motion. These can be found by setting the left hand sides  of Eqns.\ (\ref{eq:josephson1})--(\ref{eq:josephson4}) equal to zero. We shall ignore the rather specialized solutions $Z= \pm N/2 $ and $Y = \pm 1$ to Eqns.\ (\ref{eq:josephson2}) and (\ref{eq:josephson4}), and consider only the solutions $\alpha = \{0,\pi\}$, and $\beta = \{0,\pi\}$. This leads to four different combinations for $\alpha$ and $\beta$ which we can insert into the other two equations of motion, and solve numerically. Finally, the energies of the static solutions are found by plugging them into the hamiltonian (\ref{eq:mfahamiltonian}). The results of this procedure are plotted in  Fig.\ \ref{fig:loopwithtilt} as a function of the tilt $\Delta \epsilon$ (we have set $\Delta \epsilon^a=\Delta \epsilon$ for simplicity). Each panel corresponds to a different value of the boson-impurity interaction $W$, and for comparison we have also included the many-body eigenvalues for a system with $N=6$ bosons. Even for this small number of particles, the lowest and highest mean-field solutions cling closely to the lowest and highest many-body energies. The two intermediate mean-field solutions (and also the loops in the lowest and highest mean-field solutions) do not each cling to a single many-body state, but rather, they pass right through avoided crossings, jumping between two many-body states in the process. This behaviour can also be seen in Figs.\ \ref{fig:pitchforkenergy} and \ref{fig:pitchsmallN}. The avoided crossings may be viewed as a tunneling effect: when either of the relative phases $\alpha$ or $\beta$  is close to $\pi$ then we are near the barrier top of a cosine potential [see the hamiltonian Eq.\ (\ref{eq:mfahamiltonian})] and tunneling corrections to the mean-field solution become important.

The  mean-field solution with $\alpha=\beta=0$ (plotted with black dots) corresponds to both pendula pointing down, and has the lowest energy. The next solution up for our parameters has $\alpha=\pi$ and $\beta=0$  (orange squares), and corresponds to the BEC pendulum pointing down and the impurity pendulum pointing up. The solution with $\alpha=0$ and $\beta=\pi$ (blue diamonds) corresponds to the BEC pendulum pointing  up and the impurity pendulum pointing  down, and is the third most energetic solution for $\Delta \epsilon=0$ (various branches of different solutions can cross for $\Delta \epsilon \neq 0$).  The highest energy solution has $\alpha=\beta=\pi$ (red triangles), corresponding to both pendula pointing up. 

The structure of the mean-field solutions is quite rich and changes qualitatively as the parameters change. As can be seen in Fig.\ \ref{fig:loopwithtilt}, the bands can contain swallowtail loops. These are a manifestation of the nonlinearity of the mean-field theory and have been studied previously in the context of BECs in double well potentials \cite{Karkuszewski02},  in the band structure of  BECs in optical lattices \cite{wu02,machholm03,mueller02,pethick}, and also in the band structure of non-interacting atoms in optical cavities \cite{venkatesh11}. A recent experiment on a BEC trapped in a double well potential  has seen evidence for the existence of the loop structure through the violation of adiabaticity \cite{IBloch}.

Of particular significance is the fact that the lowest band (black dots in Fig.\ \ref{fig:loopwithtilt}) undergoes a bifurcation and develops a swallowtail loop as the magnitude of $W$ is increased beyond a critical value, $\pm W_{c}$ (i.e.\ for both attractive and repulsive boson-impurity interactions). This suggests that the ground state of the system changes significantly at $W_{c}$. A zoom-in of the birth of the loop at the birfurcation is shown in Fig.\ \ref{fig:groundloop}. In the top panel (a) we have $W < W_{c}$, and the band is a smooth curve. In the middle panel (b) we have $W =W_{c}$, at which point a cusp forms at zero tilt, heralding the birth of the loop. In the bottom panel (c) we have $W>W_{c}$, and the band contains a loop.   For the case of a pure BEC (no impurity) in a double well potential, a loop can also appear in the lowest band, but only when the interboson interactions are attractive ($U<0$). Here it can occur for entirely repulsive interactions (both $U>0$ and $W>0$).

Let us now focus on the lowest band, which is defined by the phase differences $\alpha=\beta=0$. The lowest band is given by the simultaneous solution of the equations
\begin{eqnarray}
\Delta \epsilon^a + 2 W Z + 4 J^a \frac{Y}{\sqrt{1-4Y^{2}}}   &=& 0 \label{eq:lowenband1} \\
	\Delta \epsilon + 2UZ + 2 W Y + 4 J \frac{Z}{\sqrt{N^2-4 Z^{2}}} &=& 0  
	\label{eq:lowenband2}
\end{eqnarray}
To remove the square roots we introduce the variables $\theta$ and $\phi$:
\begin{eqnarray}
 Y & \equiv &  \frac{1}{2}\sin{\phi} \qquad -\frac{\pi}{2}<\phi<\frac{\pi}{2} \\
  Z  & \equiv & \frac{N}{2} \sin{\theta} \qquad -\frac{\pi}{2}<\theta<\frac{\pi}{2} 
  \end{eqnarray}
Eqns.\ (\ref{eq:lowenband1}) and (\ref{eq:lowenband2}) can then be combined to give  the function 
\begin{equation}
\begin{split}
	F(\theta)  \equiv &  \Delta \epsilon + NU \sin{\theta} + 2J \tan{\theta} \\  & - W \frac{\Delta \epsilon^a+ N W \sin{\theta}}
	{\sqrt{4 {J^a}^2+\left(\Delta \epsilon^a+ N W \sin{\theta}\right)^2}} =0
	\label{eq:thetafunction}
\end{split}
\end{equation}
where we have made use of the identity $\sin{\phi}=\tan{\phi}/\sqrt{1+\tan^2\phi}$. Solutions of Eq.\ (\ref{eq:thetafunction}), i.e.\ $F(\theta)=0$, are plotted as solid curves in Fig.\ \ref{fig:groundloop}.

The lowest band typically has small population differences $Z$ between the two wells, and so we can restrict our attention to small values of $\theta$. A plot of $F(\theta)$ versus $\theta$ reveals that for small $\theta$ it is either linear, or for some parameter values it can develop a cubic structure. This fits in with what we see in Fig.\ \ref{fig:groundloop}, because there is either one or three solutions at each value of the tilt $\Delta \epsilon$, depending upon whether $W$ is less or greater than $W_{c}$. To obtain the explicit cubic equation describing the loop, we make a Taylor expansion of $F(\theta)$ up to third order about $\theta=0$ to give
\begin{widetext}
\begin{align}
	\begin{split}
		F_3(\theta) =& \frac{1}{6}\left(4J-NU + \frac{4{J^a}^2 N W^{2} \left(\left({\Delta \epsilon^a}^2 +
		4{J^a}^2\right)^2+12 N^2 W^2 \left({J^a}^2 - {\Delta \epsilon^a}^2\right)\right)}{\left({\Delta \epsilon^a}^2 + 4
		{J^a}^2\right)^{7/2}}\right) \theta^{3} \\
		 & + 6\frac{\Delta \epsilon^a {J^a}^2 N^2 W^{3}}{\left({\Delta \epsilon^a}^2 + 4{J^a}^2\right)^{5/2}} \theta^2
		+ \left(2J+NU - \frac{4{J^a}^2 N W^{2}}{\left({\Delta \epsilon^a}^2 + 4{J^a}^2\right)^{3/2}}\right) \theta 
		+ \Delta \epsilon - \frac{\Delta \epsilon^a W}{\sqrt{{\Delta \epsilon^a}^2+4{J^a}^2}} \, .
	\end{split}
	\label{eq:taylorexpansion}
\end{align}
\end{widetext}
The solutions to the equation $F_{3}(\theta)=0$ are plotted as the red dots in Fig.\ \ref{fig:groundloop}. As can be seen, there is good agreement with the solutions of the full function $F(\theta)=0$.

\begin{figure}
		\includegraphics[width=0.9\columnwidth]{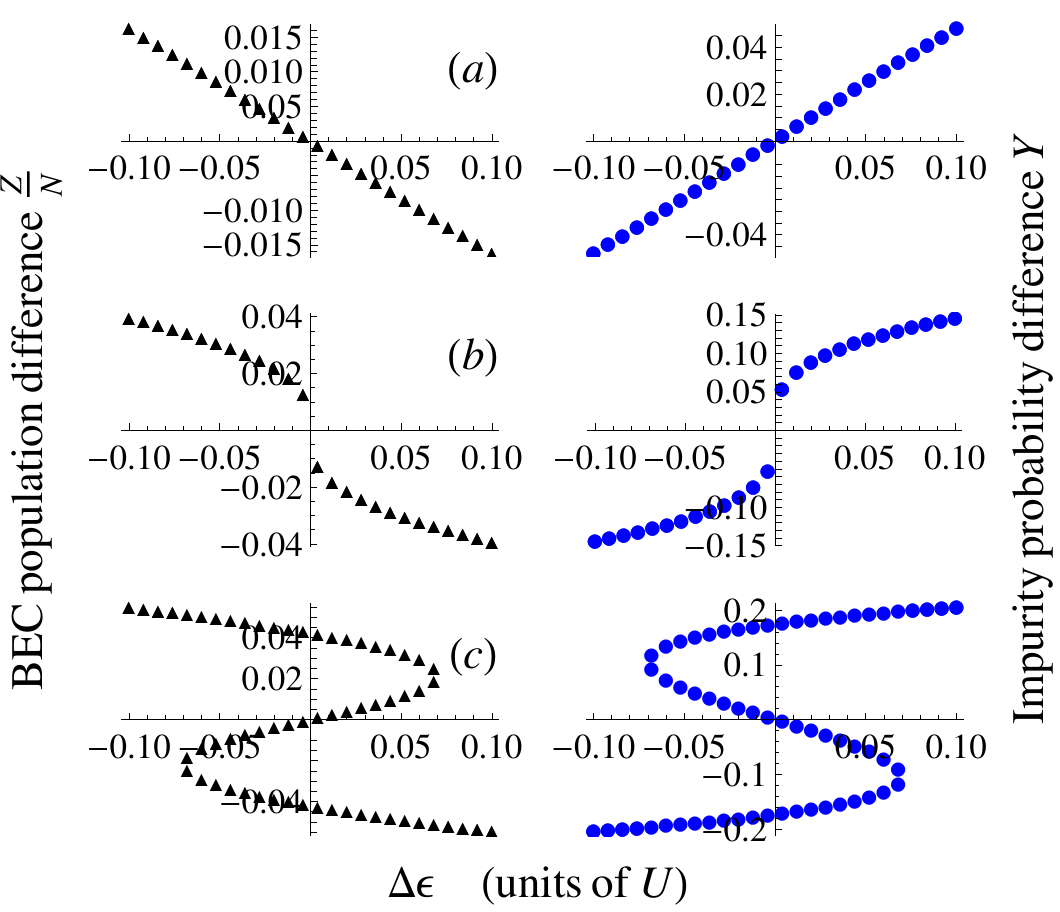}
	\caption{(Color online) The atom number difference between the right and left hand wells for the lowest band plotted as a function of the tilt $\Delta \epsilon$ ($=\Delta \epsilon^{a}$). The left hand column gives relative atom number difference $Z/N = (N_{R}-N_{L})/2N$ for the BEC, and the right hand column gives the probability difference $Y=(M_{R}-M_{L})/2$ for the impurity. Each row has a different value of the boson-impurity interaction $W$, and corresponds to the equivalent panel of Fig.\ \ref{fig:groundloop}: (a) $W/U=2.0$; (b) $W/U=2.122$; (c) W/U=2.2. The other parameters are set at $N=6$, $J=J^{a}=1.5 \, U$.}
	\label{fig:population}
\end{figure}

Let us calculate the critical value of the boson-impurity interaction $W_{c}$ at which the bifurcation in the lowest band occurs. From Fig.\ \ref{fig:groundloop}, we see that the loop is born at $\Delta \epsilon = \Delta \epsilon^{a}=0$. Furthermore, consideration of the way $F(\theta)$ evolves between a linear and a cubic function shows that the bifurcation occurs when
the first derivative of $F(\theta)$ or $F_3(\theta)$ vanishes at $\theta=0$. Thus, we find
\begin{equation}
	\frac{W_{c}}{U} = \sqrt{\frac{2J^a}{N U}\left(\frac{2 J}{U}+N \right)} \, .
	\label{eq:loopcondition}
\end{equation}
For example, when $J=J^{a}=1.5 \, U$, and $N=6$ we find $W_{c} \approx 2.121 \, U$, which is the value used in panel (b) of Figs.\ \ref{fig:groundloop} and \ref{fig:population}. 

What value does the ratio $W_{c}/U$ take in a realistic experimental situation? Let us consider the two key experiments, Gati \emph{et al} \cite{albiez05} and Levy \emph{et al} \cite{levy07}, where macroscopic tunneling and allied Josephson effects were first seen in single bosonic Josephson junctions. In the former experiment $N=1150$ and $J/N U=1/30$, and in the latter experiment $N=10^5$ and $J/NU=1/600$. Although there was no impurity present in either of these experiments, if for the sake of argument we assume that $J^{a}=J$, then we find that $W_{c}/U = 9$, and $W_{c}/U = 18$, respectively. If it doesn't occur naturally, this factor of $10$ between the boson-impurity and the boson-boson interaction energies can be achieved using a Feshbach resonance, see, for example, the experiment \cite{gross10} on the internal Josephson effect which employed a Feshbach resonance to control interactions between different hyperfine states of Bose-Einstein condensed $^{87}$Rb atoms.

\begin{figure}
		\includegraphics[width=0.9\columnwidth]{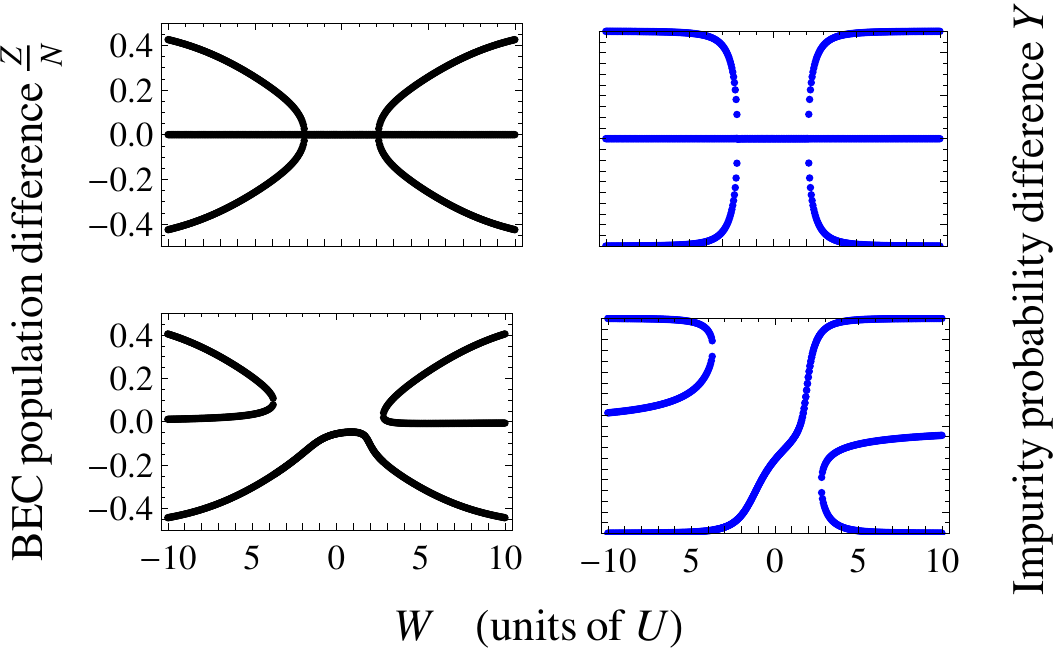}
	\caption{(Color online) The atom number difference between the right and left hand wells for the lowest band plotted as a function of the boson-impurity interaction strength  $W$. The left hand column gives relative atom number difference $Z/N = (N_{R}-N_{L})/2N$ for the BEC, and the right hand column gives the probability difference $Y=(M_{R}-M_{L})/2$ for the impurity. The top row has zero tilt: $\Delta \epsilon = \Delta \epsilon^{a} =0$, and the bottom row has a finite tilt: $\Delta \epsilon = \Delta \epsilon^{a} = U$.
 The other parameters are set at $N=6$, $J=J^{a}=1.5 \, U$.}
	\label{fig:pitchfork}
\end{figure}

The lowest band is defined by the condition that the relative phases between the two wells are zero. However, the relative number differences $Y$ and $Z$ between the two wells can vary along the band. It is particularly interesting to see how the number differences behave in the vicinity of the loop in the lowest band, and this is illustrated in Figs.\  \ref{fig:population} and {\ref{fig:pitchfork}}. Consider Fig.\ \ref{fig:population} first, which shows how $Z$ and $Y$ vary with the tilt $\Delta \epsilon$. The top panel (a) has $W<W_{c}$, and there is only a single solution for $Y$ and $Z$ at each value of $\Delta \epsilon$, as expected. Because we have chosen a positive value of $W$, which corresponds to repulsive boson-impurity interactions, the BEC and the impurity have opposite dependences upon $\Delta \epsilon$ (we have set $\Delta \epsilon^{a}=\Delta \epsilon$ for simplicity). For our parameters, the BEC  has the greater probability of occupying the lower well, so that $Z \equiv (N_{R}-N_{L})/2$ is positive when $\Delta \epsilon \equiv \epsilon_{R}-\epsilon_{L}$ is negative. Conversely, the impurity has the greater probability of occupying the upper well, and so $Y \equiv (M_{R}-M_{L})/2$ is negative when $\Delta \epsilon$ is negative. The middle panel (b) has $W=W_{c}$, and features a very sudden change in $Y$ and $Z$ at zero tilt. Finally, the bottom panel (c) has $W > W_{c}$ and the population difference has developed a portion which is folded back to give the classic ``S'' shape associated with a fold catastrophe \cite{venkatesh11}. This latter structure can generally be expected to lead to a hysteresis effect when the tilt is swept through zero in one direction versus the other direction. Consider, for example, the case when $W>W_{c}$, and $\Delta \epsilon$ is large and negative. Then $Z$ will begin on its upper branch and $Y$ on its lower branch. Increasing the tilt, both will follow their respective branches until each branch vanishes (i.e. each curve folds back to form the middle branches) at a finite positive value of the tilt. The system is then forced to make a jump to the lower and upper branches, respectively (or even to other bands). Conversely, if the sweep is performed from positive to negative value of the tilt the jump will occur at a finite negative value of the tilt.  This sudden disruption to the evolution of the populations as the tilt is changed means that adiabatic evolution is impossible in the presence of loops \cite{wu02,venkatesh11,IBloch,wu00}, at least in the mean-field approximation.

Whereas Fig.\ \ref{fig:population} shows the dependence of the population difference upon the tilt, Fig.\ \ref{fig:pitchfork} shows its dependence upon the boson-impurity interaction $W$. We see that the latter situation gives rise to a pitchfork bifurcation, as expected from the cubic form of $F(\theta)$. In fact, because Fig.\ \ref{fig:pitchfork} shows a range of $W$ that extends from positive to negative values, we find back-to-back pitchfork bifurcations due to the fact that there are bifurcations at $\pm W_{c}$. The top row of Fig.\ \ref{fig:pitchfork} is for zero tilt, and gives a symmetric pitchfork, whereas the bottom row is for a finite value of the tilt and gives a broken pitchfork. The bottom row shows that there can be qualitative differences between the behavior of the BEC and the impurity as $W$ is swept through zero (if the tilt is non-zero). For example, if the BEC starts off on the lowest branch for $W < 0$, and $W$ is swept through to $W>0$, it remains on the lowest branch (adiabatic evolution) and $Z$ is negative throughout, i.e.\ the BEC remains in the lower-lying left well. Meanwhile, if the impurity also starts off on its lowest branch for $W<0$ (as it very well might do if the BEC starts off on its lowest branch, because then it is in contact with the BEC, thereby lowering the energy), and $W$ is then swept to positive values, we find that the impurity can also evolve adiabatically, but is transferred into the higher lying right well by the end of the sweep.

The top row of panels in Fig.\ \ref{fig:pitchfork}, which is for zero tilt, is perhaps even more interesting because it illustrates a spontaneous breaking of the left/right symmetry of the number differences $Y$ and $Z$ when $ \vert W \vert >W_{c}$.
In the many-body case the system can in principle be in a superposition of more than one branch of the pitchfork, but in the mean-field description it must choose just one value for each of the number differences. Note that when $\vert W \vert < W_{c}$ we expect both $Y$ and $Z$ to occupy the equivalent branch of their pitchforks, because that lowers the energy. Conversely, when $\vert W \vert > W_{c}$ they will occupy opposing branches, because again this lowers the energy. The middle branch is expected to be unstable. 

A pitchfork bifurcation of the atom number difference between two modes in a BEC has recently been experimentally investigated by the Oberthaler group \cite{zibold2010}. In their case, the bifurcation was at the transition between the Rabi and Josephson regimes (see Section \ref{sec:coherence}) of the internal Josephson effect where, rather than a physical double well potential, the two modes are provided by two spin states.

In Fig.\ \ref{fig:pitchforkenergy} we have plotted the many-body and mean-field energies as a function of the boson-impurity interaction $W$. To see how this figure fits in with the figures showing the swallowtail loops, imagine adding a third axis to Fig.\ \ref{fig:loopwithtilt}, that comes out of the page and along which $W$ is increased. Then Fig.\  \ref{fig:pitchforkenergy} shows a slice through this 3D space for a fixed value of $\Delta \epsilon (= \Delta \epsilon^{a})$. The loop in the lowest band (black circles) now presents itself as a pitchfork bifurcation (which is broken due to the finite value of  $\Delta \epsilon$ chosen in Fig.\ \ref{fig:pitchforkenergy}). The smaller dots are the other mean-field bands (the color and symbol scheme follows that of Fig.\ \ref{fig:loopwithtilt}). For comparison we have also included the many-body eigenvalues (solid curves). 
To avoid confusion we remind the reader that the hamiltonian (\ref{eq:twomodehamiltonian}) neglects constant factors that only depend on the total number of atoms. This explains why the ground state energy in Fig.\ \ref{fig:pitchforkenergy} decreases as $W$ increases. The true boson-impurity interaction energy $(N+\Delta N \Delta M)W/2$ tends to zero for large $W$, but here we have dropped the $N W/2$ part.

As can be seen in Fig.\ \ref{fig:pitchforkenergy}, the mean-field static solution we have been referring to as the ``lowest band'' (defined by $\alpha=\beta=0$) is not always the lowest in energy: for certain values of $W$  branches of the solution with $\alpha=0$ and $\beta=\pi$ (blue dots) can have a lower energy than the top and middle branches of the $\alpha=\beta=0$  solution. Nevertheless, the lower branch of the  $\alpha=\beta=0$ solution still has the lowest energy, and clings very closely to the many-body ground state energy.

It is notable that the middle branch of the pitchfork in Fig.\ \ref{fig:pitchforkenergy} forms a lower envelope bounding a region of avoided crossings (with the exception of the two small regions where the $\alpha=0$, $\beta= \pi$ band pierces the pitchfork). This is in accordance with studies \cite{Karkuszewski02} of BECs in double well potentials with no impurity, where the mean-field solutions bound regions of avoided crossings.

In order to gain a detailed understanding of the behavior of the system as a function of $W$, we have plotted the $N=6$ case in Fig.\ \ref{fig:pitchsmallN}. We have reduced the value of $J/U$ from that used in Fig.\  \ref{fig:pitchforkenergy} so that the eigenstates of the hamiltonian correspond quite closely to the eigenstates of the number difference operators $\Delta \hat{N}$ and $\Delta \hat{M}$ except close to the avoided crossings (note that \emph{all} apparent crossings are in fact avoided crossings), which makes the picture easier to interpret. For large values of $W$ the energies are dominated by the $(W/2)\Delta N \Delta M$ boson-impurity interaction. This gives straight lines as a function of $W$ whose gradients are  $\Delta N \Delta M /2$, and this fact allows us to identify which asymptotic state is which. When $N=6$ we only have the possibilities $\Delta N = \pm 6, \pm 4, \pm 2, 0$, and $\Delta M= \pm 1$. This is what we see in Fig.\ \ref{fig:pitchsmallN}, where pairs of states can be identified at large $W$ with gradients $-3W$, $-2W$, $-W$, and zero (we only show the lower portion of the energy space and so the states with positive gradients at large $W$ are not shown). Consider the two number states with gradient $-3W$; the lower energy state of the pair has all six bosons in the lower (left) well  so that $\Delta N=-6$ and the impurity in the higher well so that $\Delta M=1$. The higher energy state of the pair has the converse.  For large enough $W$ these two states become the ground state and the first excited state. Fig.\  \ref{fig:pitchsmallN} allows us to connect mean-field solutions to many-body states: the lowest branch of the pitchfork accurately gives the ground state energy of the hamiltonian for all $W$; the middle branch of the pitchfork gives the number state which has the same magnitudes of $\Delta N$ and $\Delta M$ as the ground state, but with reversed signs, and which for large $W$ (i.e.\ beyond the last avoided crossing at $W \approx 33 \, U$) becomes the first excited state of the hamiltonian, and the top branch gives the number state with $\Delta N=0$ and $\Delta M=-1$. When the tilt is zero the middle and lower branches of the pitchfork become degenerate with each other.

\section{Many-body eigenstates}
\label{sec:eigenstates}

\begin{figure}
\includegraphics[width=0.9\columnwidth]{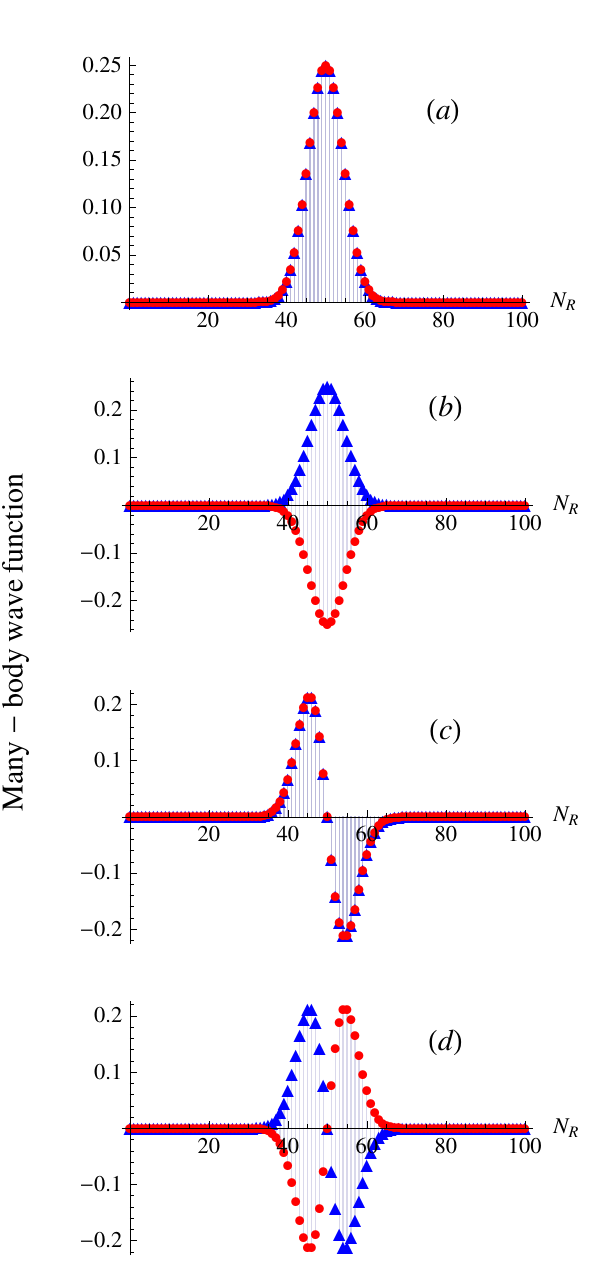}
\caption{(Color online) The first four many-body eigenstates for $W=0$. The ground state is shown in panel (a), the first excited state in panel (b) and so on. The red circles/blue triangles are the amplitudes for the impurity to be in the left/right well.
The total number of atoms is $N=100$ and $N_{R}$ is the number of atoms in the right well. The hopping energy is set to $J=J^{a}=10 \, U$ and the tilt is zero. According to hamiltonian (\ref{eq:twomodehamiltonian}), the energies of these four states are $-0.0995, -0.0975, -0.0947$, and $-0.0927$ in units of $UN^2$.}
	\label{fig:evecsig0}
\end{figure}

\begin{figure}
\includegraphics[width=0.9\columnwidth]{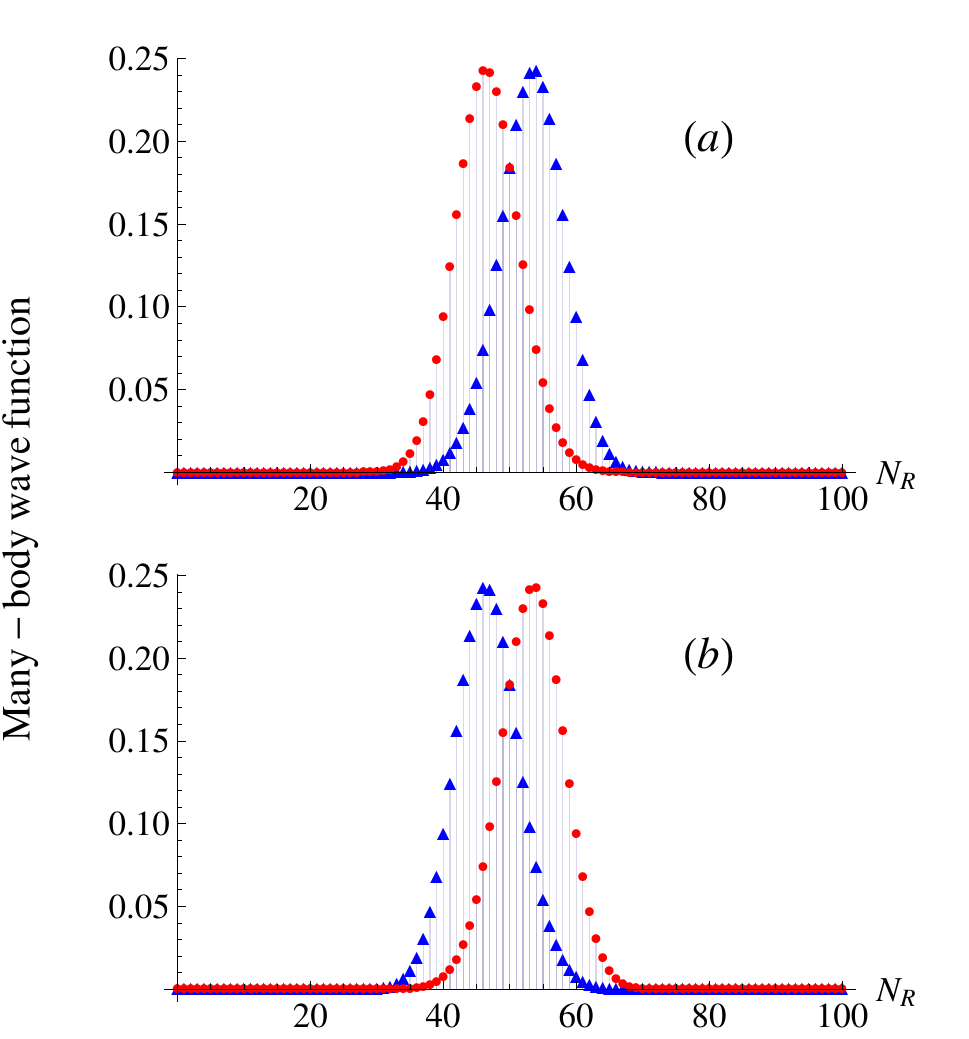}
\caption{(Color online) The ground state for $W/U=-10 $ (a), and $W/U=10 $ (b). The red circles/blue triangles are the amplitudes for the impurity to be in the left/right well.
The total number of atoms is $N=100$ and $N_{R}$ is the number of atoms in the right well. The hopping energy is set to $J=J^{a}=10 \, U$ and the tilt is zero. According to hamiltonian (\ref{eq:twomodehamiltonian}), both the positive and negative values of $W$ lead to the same ground state energy of $-0.101$ in units of $UN^2$.}
	\label{fig:evecsig10}
\end{figure}

The many-body eigenstates can be written as superpositions of Fock states  $\vert M_{R},N_{R} \rangle$
\begin{equation}
\vert \Psi^{j} \rangle = \sum_{M_{R},N_{R}} C^{j}_{M_{R},N_{R}} \vert M_{R},N_{R} \rangle \, ,
\label{eq:manybodyeigenstate}
\end{equation}
where the index $j$ labels the $j^{\mathrm{th}}$ eigenstate.
For every possible value of $N_{R}$ for the bosons, there are two possible values of $M_{R}$, corresponding to the impurity being in either the left or the right well.
Some examples of these eigenstates are shown in Figs.\ \ref{fig:evecsig0} and \ref{fig:evecsig10}. The red circles correspond to the impurity being in the left well, i.e.\ $C_{0,N_{R}}$, and the blue triangles to it being in the right well, i.e.\ $C_{1,N_{R}}$. In Fig.\ \ref{fig:evecsig0} the boson-impurity interaction $W=0$ and so in the ground state shown in panel (A) the circles and triangles sit on top of each other. 
Although the state is peaked around $\Delta N =0$, where $\Delta N = N_{R}-N_{L}=2N_{R}-N$, the many-body state has a finite width in $\Delta N$, unlike the mean-field state. This width can be estimated by observing that the low-lying states of the pendulum model discussed in Section \ref{sec:mfa} see an essentially harmonic potential. Applying the results of the quantum harmonic oscillator gives the ground state wave function (for either one of the impurity states, and assuming $W=0$) as
\begin{equation}
\Psi_{0}(Z)=\frac{1}{\pi^{1/4}} \left({\frac{2U}{JN}}\right)^{1/8} \exp \left[-\frac{Z^2}{2}\sqrt{\frac{2U}{JN}} \right]
\end{equation}
where $Z = \Delta N/2$. In the case of a finite number of bosons this wave function is sampled discretely, but nevertheless it gives a good approximation to the exact result shown in Fig.\ \ref{fig:evecsig0} (A) when $N$ is large. In particular, the width varies as $(JN/2U)^{1/4}$. In the interaction dominated regime $J \ll U$, the probability distribution for the ground state becomes narrow and tends to a single Fock state for which there are exactly $N/2$ bosons in each well.

The symmetry of the hamiltonian (\ref{eq:twomodehamiltonian}) means that its eigenstates have definite parity.
The ground state shown in  Fig.\ \ref{fig:evecsig0} (a) is an even function of the impurity location, whereas the first excited state shown in panel (b) is an odd function of the impurity location. The second excited state is shown in panel (c) and this is even in the impurity degrees of freedom but odd in the bosonic degrees of freedom specified by $N_{R}$, the number of bosons in the right well (the parity of the bosonic degrees of freedom is defined with respect to $\Delta N$). Finally, the third excited state is shown in panel (d) and this is odd in both the impurity and bosonic degrees of freedom.

In Fig.\ \ref{fig:evecsig10} we show just the ground state, but this time for two different non-zero values of $W$; panel (a) has $W/U=-10 $ and panel (b) has $W/U=10$. In both cases the wave function has been split by the interaction with the impurity, but notice that the circles and the triangles swap positions between the two panels. In the repulsive case the system lowers its energy by placing more bosons in the opposite well to the impurity. In the limit $W \rightarrow \infty$ the impurity will occupy one well and all the bosons the other, but when the tilt is zero there is nothing to decide between the two wells and so the system enters a balanced macroscopic superposition of being in both wells. We then expect the ground state to take a  Sch\"{o}dinger cat form $\vert \Psi^{0} \rangle =  (\vert 1, 0 \rangle + \vert 0,N \rangle)/\sqrt{2}$.
In the attractive case the system lowers its energy by placing more bosons in the same well as the impurity,  
and again the system is forced into a macroscopic superposition of occupying both wells \cite{spekkens99}. As $W \rightarrow - \infty$ we expect the ground state of the system to be $\vert \Psi^{0} \rangle =  (\vert 1, N \rangle + \vert 0,0 \rangle)\sqrt{2}$. We shall discuss Schr\"{o}dinger cat states further in Section \ref{sec:schrodingercat}.

\begin{figure}
\includegraphics[width=0.9\columnwidth]{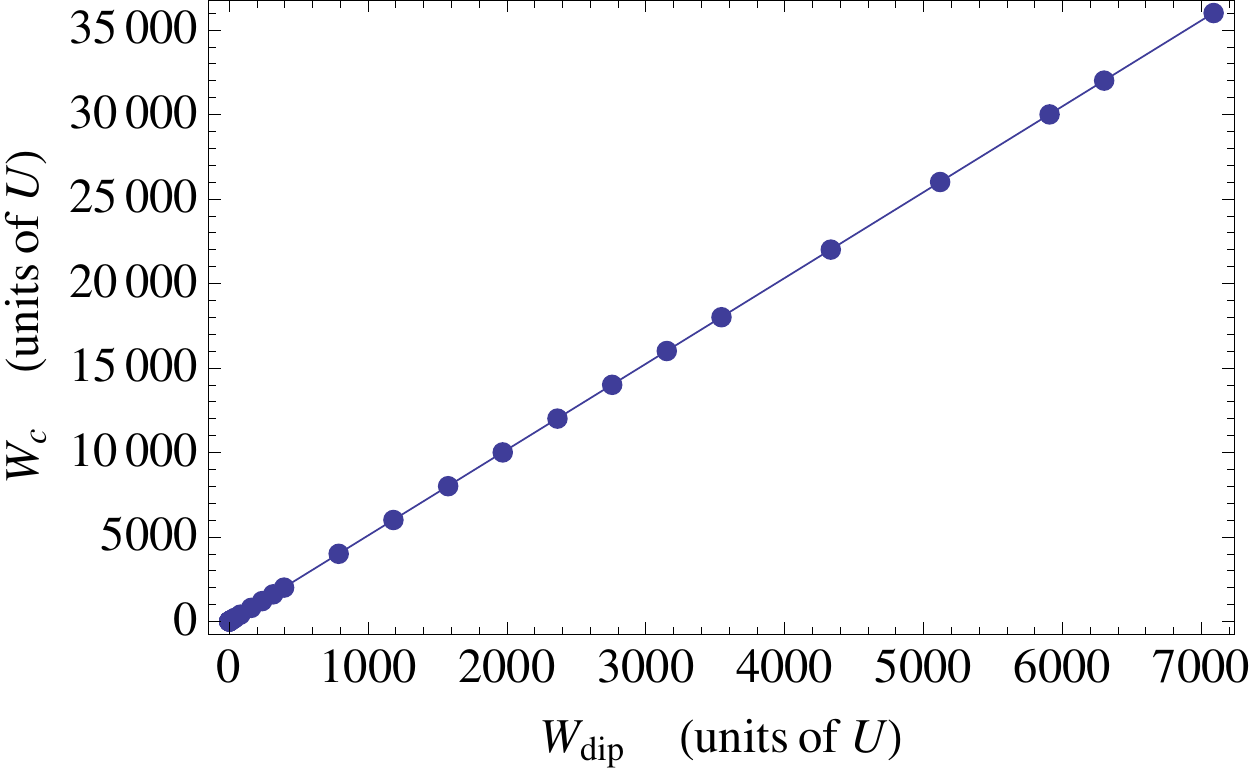}
\caption{(Color online) The correlation between critical value $W_{c}$ [see Eq.\ (\ref{eq:loopcondition})] of the boson-impurity interaction at which a bifurcation occurs in the lowest band, and the equivalent value $W_{\mathrm{dip}}$  at which the many-body wave function first develops a dip in its center (i.e.\ at $N_{R}=N/2$). Each data point corresponds to $N=100$, but to a different value of $J=J^{a}$, lying in the range $1<J/U<180,000$. The solid line is a fit to the data and is given by  $W_{c}=-3.8+5.1 W_{\mathrm{dip}}.$}
	\label{fig:cuspvsdip}
\end{figure}

An intriguing question concerns whether there is connection between the loop bifurcation in the mean-field solutions and the splitting of the many-body quantum state? In fact there is a direct connection, as indicated by Fig.\ \ref{fig:cuspvsdip}. This figure plots $W_{c}$ versus $W_{\mathrm{dip}}$, where $W_{c}$ is the mean-field prediction, given by Eq.\ (\ref{eq:loopcondition}), for the critical value of $W$ at which the bifurcation in the lowest band occurs, and $W_{\mathrm{dip}}$ is the value of $W$  at which the ground state probability distribution for $N_{R}$ first develops a dip at $N_{R}=N/2$, heralding the onset of the splitting. Each point in Fig.\ \ref{fig:cuspvsdip} corresponds to a different value of the hopping energy $J=J^{a}$, which lies in the range $1<J/U<180,000$. Performing a linear fit to the data points shown in Fig.\ \ref{fig:cuspvsdip}, we find that $W_{c} = -3.8+  5.1 \, W_{\mathrm{dip}}$, indicating that the many-body wave function begins splitting before the mean-field bifurcation. However, the splitting of the many-body wave function is a smooth process (for finite $N$) and there is no particular point at which we can say the wave function has split. We have chosen $W_{\mathrm{dip}}$ as a simple, but arbitrary indicator. Another choice might be when the two peaks of the many-body wave function are `resolvable', i.e.\ separated by the width of the individual gaussians. With this indicator (not shown) we find the many-body wave function splits after the mean-field bifurcation. Nevertheless, it is clear from Fig.\ \ref{fig:cuspvsdip} that whatever the precise choice of indicator  there is a direct correlation between the occurrence of the mean-field loop and the splitting of the many-body wave function.

\section{Coherence versus number fluctuations in the ground state}
\label{sec:coherence}

Following Gati and Oberthaler \cite{gati07}, let us calculate the coherence and number fluctuations of the ground state. The coherence is related to the visibility of the interference fringes which are formed if the atoms are released from the double well potential, and the atomic clouds allowed to ballistically expand, until they overlap and are imaged in a Young's double slit type experiment. For the bosons the coherence is defined to be
\begin{equation}
\alpha \equiv \langle \Psi \vert \hat{B} \vert \Psi \rangle/N 
\end{equation} 
where $\hat{B}=\hat{b}_{L}^{\dag}\hat{b}_{R}+\hat{b}_{R}^{\dag}\hat{b}_{L}$ is the bosonic hopping operator. 
In terms of the amplitudes $C_{M_{R},N_{R}}$ defined in Eq.\ (\ref{eq:manybodyeigenstate}), one finds that
\begin{equation}
\begin{split}
\alpha =& \frac{1}{N} \sum_{M_{R},N_{R}} C_{M_{R},N_{R}-1}^*C_{M_{R},N_{R}}\sqrt{N_{R}(N-N_{R}+1)} \\
	&+ C_{M_{R},N_{R}+1}^*C_{M_{R},N_{R}}	\sqrt{(N-N_{R})(N_{R}+1)} \\
	=& \frac{1}{N}\sum_{N_{R}} \left(C_{0,N_{R}-1}^*C_{0,N_{R}} + C_{1,N_{R}-1}^*C_{1,N_{R}} \right) \\
	& \quad \times \sqrt{N_{R}(N-N_{R}+1)} \\
	&+ \left(C_{0,N_{R}+1}^*C_{0,N_{R}} + C_{1,N_{R}+1}^*C_{1,N_{R}} \right) \\
	& \quad \times \sqrt{(N-N_{R})(N_{R}+1)} \, . \label{eq:coherence}
\end{split}
\end{equation}
The coherence is therefore large when neighboring amplitudes in Fock space are large, i.e.\ when the number distribution is made up of a single broad peak. On the contrary, it is small if the number distribution consists of individual spikes which are separated by at least $\delta N_{R}= 2$. The two square root factors are peaked just either side of $N_{R}=N/2$, and so favor a number distribution which is centered around $N_{R}=N/2$. 
The above result will only give the visibility of the boson fringes if the state of the impurity is not measured (a modified result is obtained if the location of the impurity is known).

\begin{figure}
\includegraphics[width=0.9\columnwidth]{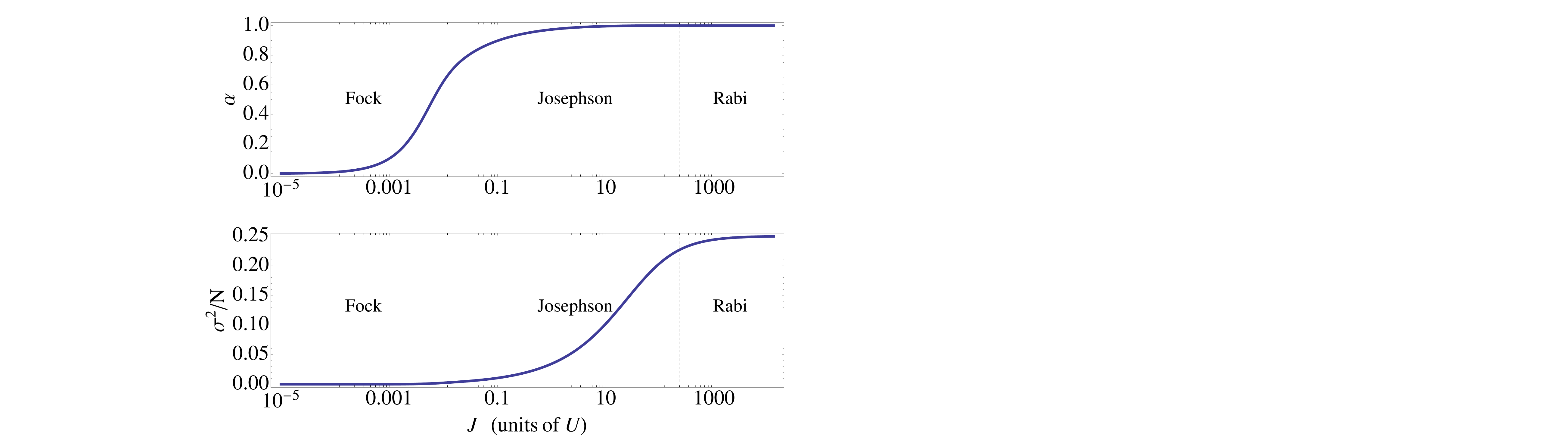}
\caption{(Color online) The coherence $\alpha$, and the square of the number fluctuations $\sigma^2$, in the ground state as a function of the boson hopping energy $J$, with $N=100$. The boson-impurity interaction $W$ is set to zero, as is the tilt. The dashed lines indicate the boundaries between the Fock, Josephson, and Rabi regimes.}
	\label{fig:coherence}
\end{figure}

If, rather than a Young's double slit type measurement, an \emph{in situ} measurement of the atom number in one of the wells is made, then a single measurement will yield a single number (we assume here that the resolution of the measurement is at the single atom level). However, repeated measurements on an identically prepared system will yield different atom numbers according to the probability distribution obtained by squaring the amplitudes plotted in Figs.\ \ref{fig:evecsig0} and \ref{fig:evecsig10}. The width of the atom number probability distribution is characterised by the standard deviation $\sigma$, which we define via
\begin{equation}
\begin{split}
	\sigma^2 \equiv & \langle \hat N_R^2 \rangle-\langle \hat N_R \rangle^2 \\
	 = & \sum_{M_{R},N_{R}} |C_{M_{R},N_{R}}|^2 N_{R}^2  
	 -\left(\sum_{M_{R},N_{R}}|C_{M_{R},N_{R}}|^2 N_{R}\right)^2 \\
	=&\sum_{N_{R}} \left(|C_{0,N_{R}}|^2+|C_{1,N_{R}}|^2\right)N_{R}^2  \\ 
	& - \left(\sum_{N_{R}}
	\left(|C_{0,N_{R}}|^2+|C_{1,N_{R}}|^2\right)N_{R} \right)^2 
\end{split}
\end{equation}

In Fig.\ \ref{fig:coherence} we plot the coherence and the number fluctuations for the case when the impurity is decoupled from the bosons ($W=0$). This picture essentially repeats Fig.\ 1 of \cite{gati07}. We have indicated with dashed lines the boundaries between three commonly defined regimes \cite{leggett01}:
\begin{enumerate}
\item Rabi regime  $J/U \gg 2N$
\item Josephson regime $2N  \gg J/U \gg  2/N$
\item Fock regime $J/U \ll 2/N$.
\end{enumerate}
In the Fock regime the interaction energy dominates the hopping energy and so the number fluctuations are small and the coherence low. In the Rabi regime  the opposite is true, and the Josephson regime sits between the two. For example, Fig.\ \ref{fig:evecsig0} has $J/U=10 $, and $N=100$, which means that this figure corresponds to the Josephson regime. Note that when $J \rightarrow \infty$, i.e. the deep Rabi regime, then 
$\sigma \rightarrow \sqrt{N}/2$, as can be seen in Fig.\ \ref{fig:coherence}.

We now switch on the BEC-impurity interaction and ask whether a single impurity can affect the coherence properties of the bosons? This question is motivated by several previous theoretical studies \cite{bausmerth07,zapata03}. In \cite{zapata03} the authors consider the dynamics of a system comprising of two initially independent BECs which are then put into direct contact. They predicted the surprising result that macroscopic phase coherence is established between the two BECs as soon as a few atoms are exchanged. Likewise, in \cite{bausmerth07} a system comprising an atomic quantum dot sandwiched between two weakly coupled BECs was analyzed, and it was found that the dot was capable of mediating large amplitude Josephson-like oscillations of the particle imbalance despite the fact that the dot could only host a single atom at a time.

\begin{figure}
\includegraphics[width=0.9\columnwidth]{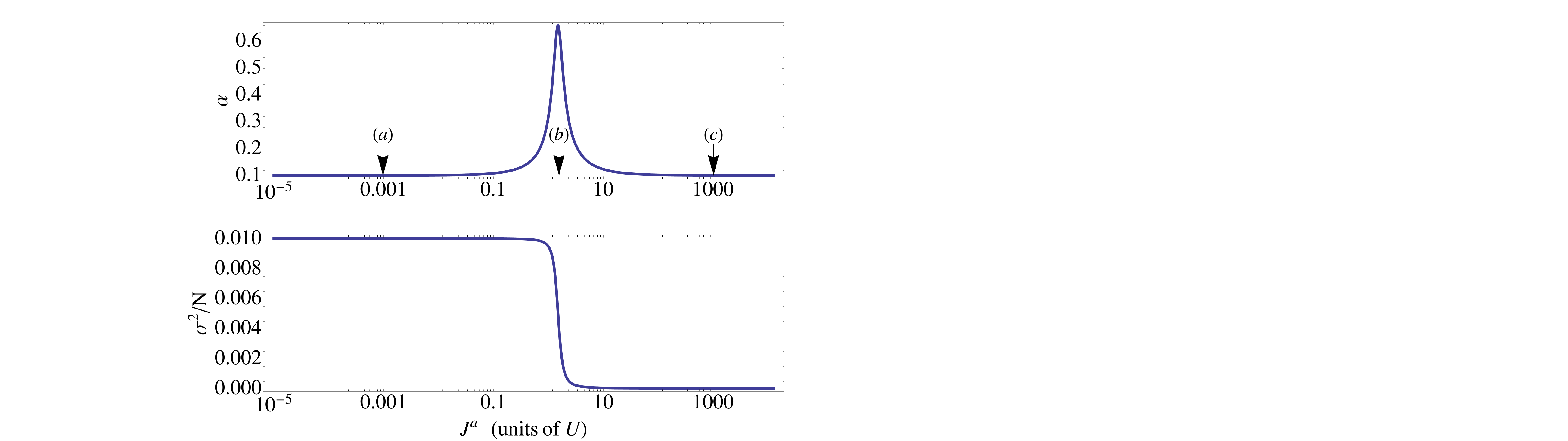}
\caption{(Color online) The coherence $\alpha$, and the square of the number fluctuations $\sigma^2$, in the ground state as a function of the impurity hopping energy $J^{a}$. In this figure $N=100$, J/U=0.001, $W/U=2 $, and the tilt is set to zero. The three labels indicate the values of $J^{a}$ used in the different panels of Fig.\ \ref{fig:eveccoherence}: (a) $J^{a}/U=0.001 $, (b) $J^{a}/U=1.514 $, and (c) $J^{a}/U=1000 $.}
	\label{fig:coherence2}
\end{figure}

 The BEC-impurity system is an ideal one in which to study these coherence-generating mechanisms because, in principle, the tunnelling properties of the impurity and the bosons can be separately controlled. For example, one can examine what happens when $J$ is very small but $J^{a}$ is not, so that the bosons are essentially frozen and only the impurity is mobile, thereby isolating its effect on the coherence between the two wells. The results are plotted in Fig.\ \ref{fig:coherence2}. In this figure the value of the boson hopping energy is $J/U=0.001$, and $N=100$, which puts the bosons in the Fock regime, and we have set $W/U= 2 $. When $J^{a}=0$ the coherence is only $\alpha =0.1$, but as $J^{a}$ is increased it peaks at over six times this value, with a maximum value of  $\alpha \approx 0.66 $  at $J^{a}/U \approx 1.5 $, which is indicated by the arrow labelled by (b) on the figure, before decaying back to the background value of $\alpha=0.1$ again as $J^{a}$ is increased further. Note that, according to Eq.\ (\ref{eq:loopcondition}), the mean-field prediction for the value of $J^{a}$ at which the bifurcation to form a loop in the lowest band occurs (when $W_{c}=2 \, U $) is $J^{a} = 2 \, U$, which is close to the peak in $\alpha$.

\begin{figure}
\includegraphics[width=0.9\columnwidth]{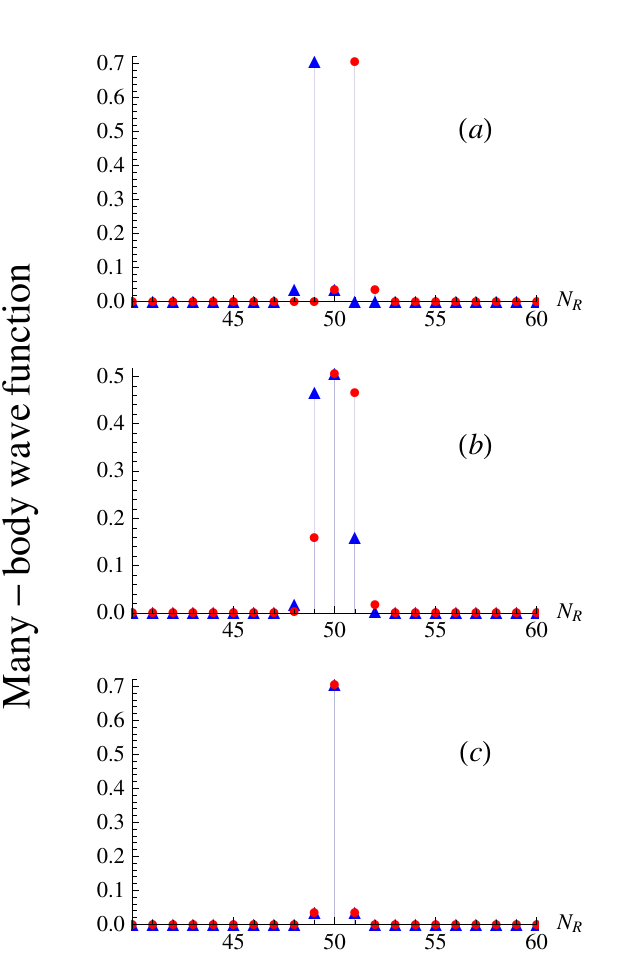}
\caption{(Color online) The many-body ground state wave function as a function of $N_{R}$, the number of atoms in the right well,  for $N=100$, $J/U=0.001$, and $W/U=2$. Each panel has a different value of the impurity hopping energy $J^{a}$ corresponding to the labels shown in Fig.\ \ref{fig:coherence2}: (a)  $J^{a}/U=0.001 $, (b) $J^{a}/U=1.514 $, and (c) $J^{a}/U=1000 $.
 The red circles/blue triangles are the amplitudes for the impurity to be in the left/right.}
	\label{fig:eveccoherence}
\end{figure}

To help us interpret this result we have plotted in Fig.\ \ref{fig:eveccoherence} the many-body wave functions corresponding to the points labelled (a), (b), and (c) in Fig.\ \ref{fig:coherence2}. By virtue of the very small value of $J/U =0.001$, the width $(JN/2U)^{1/4} \approx 0.5$ of the bosonic many-body wave function (for either impurity state) is small and we have only plotted a limited range of $N_{R}$ for clarity. The narrowness of the many-body wave function implies that the number fluctuations and coherence should be small, and this is the case \emph{except} at the peak of $\alpha$ discussed above. Let us begin with panel (c), which corresponds to the case where the many-body wave function is not split (and the mean-field loop is not present in the lowest band) because $J^{a}/U=1000$ is so large that $W_{c}$ is far above the actual value of $W=2$. The effect of the impurity is negligible in this case, and indeed the coherence and number fluctuations are small. 

Jumping now to panel (a), the wave function is split (and the loop is present) because $J^{a}/U=0.001$ is so small that $W_{c}$ is far below the actual value of $W=2$. The many-body wave function in this panel is to a good approximation given by a superposition of just two Fock states $\vert \Psi^{0} \rangle \approx (\vert 0,51\rangle+\vert 1,49\rangle)/\sqrt{2}$. Despite the fact that the total wave function is approximately two times wider than in panel (c), the coherence corresponding to (a) and (c) is the same, i.e.\ $\alpha = 0.1$. Thus, even though the wave function in (a)  has a significant total width in Fock space, the coherence is insensitive to this because each piece corresponding, respectively, to $M_{R}=1$ and $M_{R}=0$, is very narrow (and we note from Eq.\ (\ref{eq:coherence}) that the coherence is determined separately by the two pieces of the wave function). Therefore, a number measurement is  superior to a coherence measurement as a method for detecting the splitting because it can distinguish between (a) and (c). In (a) we would almost always obtain one of the values $N_{R}=49$ or $N_{R}=51$ (with equal probability), but almost never $N_{R}=50$, whereas in (c) we would almost always obtain $N_{R}=50$.

Panel (b) corresponds to the peak of $\alpha$, and shows the wave function in the process of splitting. As can be seen, the parts of the wave function giving the impurity in the left and right wells are themselves much wider than in panels (a) or (c), and hence the coherence is large. In panel (b) we therefore have a coherent state, whereas in panel (c) we have (approximately) a single Fock state and in panel (a) we have a superposition of two Fock states.
It is notable that the peak in the coherence occurs at the same value of $J^{a}$ as a step change in the number fluctuations $\sigma$ of the bosons (see lower panel of Fig.\ \ref{fig:coherence2}). As the ratio $J^{a}/U$ is increased and the impurity becomes mobile, it precipitates a sudden decrease in  $\sigma$ at the wave function splitting point.

\section{Towards Schr\"{o}dinger cat states}
\label{sec:schrodingercat}

\begin{figure}
\includegraphics[width=0.9\columnwidth]{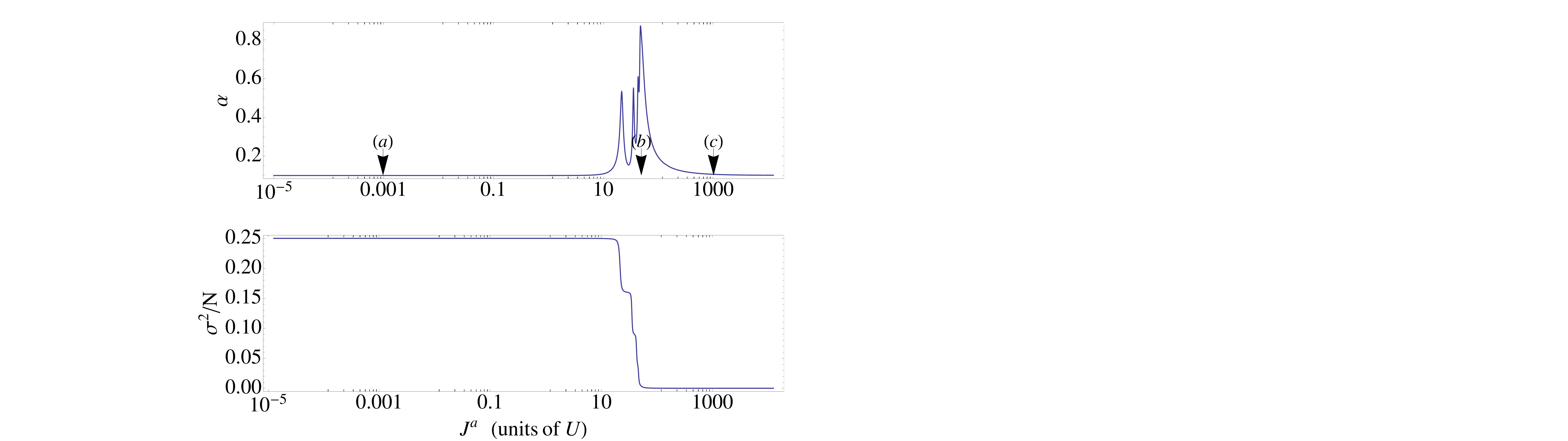}
\caption{(Color online) The coherence $\alpha$, and the square of the number fluctuations $\sigma^2$, in the ground state as a function of the impurity hopping energy $J^{a}$. In this figure $N=100$, $J/U=0.001$,  $W/U=10 $, and the tilt is set to zero. The three labels indicate the values of $J^{a}$ used in the different panels of Fig.\ \ref{fig:eveccoherence2}: (a) $J^{a}/U=0.001 $, (b) $J^{a}/U=48 $, and (c) $J^{a}/U=1000 $. A zoom-in of this figure is given in Fig.\ \ref{fig:coherence4}.}
	\label{fig:coherence3}
\end{figure}

\begin{figure}
\includegraphics[width=0.9\columnwidth]{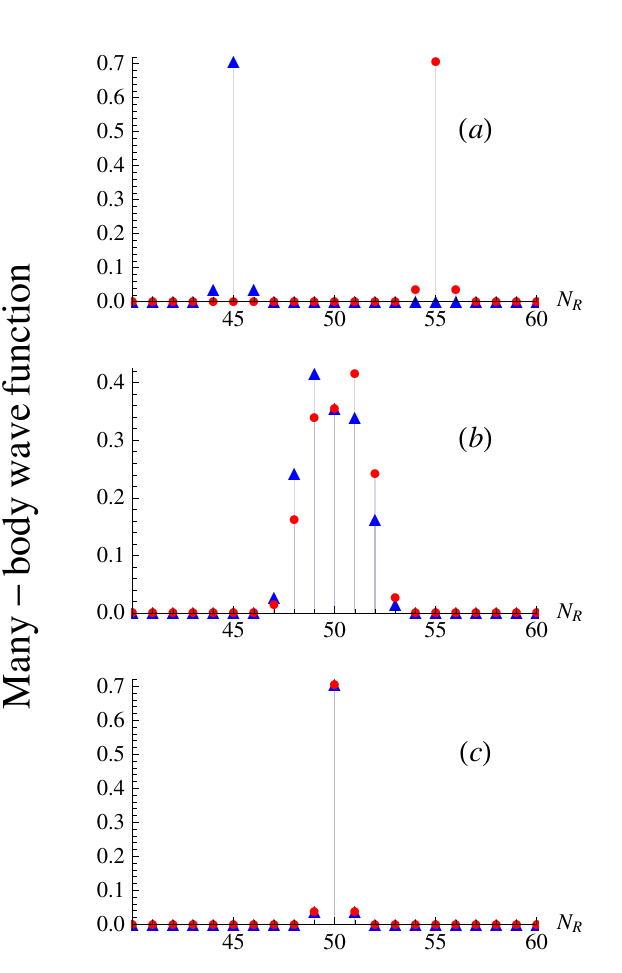}
\caption{(Color online) The many-body ground state wave function as a function of $N_{R}$, the number of atoms in the right well,  for $N=100$, $J/U=0.001$, and $W/U=10$. Each panel has a different value of the impurity hopping energy $J^{a}$ corresponding to the labels shown in Fig.\  \ref{fig:coherence3} : (a)  $J^{a}/U=0.001 $, (b) $J^{a}/U=48 $, and (c) $J^{a}/U=1000 $.
 The red circles/blue triangles are the amplitudes for the impurity to be in the left/right well.}
	\label{fig:eveccoherence2}
\end{figure}

The value of the boson-impurity interaction which we have used to illustrate the coherence properties in the previous section, namely $W= 2 \, U$, is twice the magnitude of the individual boson-boson interaction. Examining the hamiltonian (\ref{eq:twomodehamiltonian}), and neglecting the hopping operators (this is valid in the combined Fock regime when $J/U$ and $J^{a}/U$ are much smaller than unity), we find that the Fock states with $\Delta M = \pm 1$ and $\Delta N = \mp 2$ are lower in energy than the Fock state with  $\Delta M= \Delta N =0$. Increasing $\vert \Delta N \vert $ beyond $\vert \Delta N \vert = 2$ confers no further energetic advantage when $W= 2 \, U$.  This is confirmed by Fig.\ \ref{fig:eveccoherence} (a), which is in the Fock regime, where we see that the two $\delta$-function-like spikes are separated by a distance $\Delta N=2$. 

The state shown in Fig.\ \ref{fig:eveccoherence} (a) can hardly be called a Schr\"{o}dinger cat state because it is not a superposition of two macroscopically distinguishable states. Rather, it is a superposition of two states differing by a single boson. Indeed, this situation is rather similar to the case of an odd number of bosons (but no impurity), where in the Fock regime the number distribution must also be split into two pieces differing by a single particle.  In order to generate Schr\"{o}dinger cat states we must increase the value of $W$: Figs.\ \ref{fig:coherence3} and \ref{fig:eveccoherence2} illustrate the situation when $W= 10 \, U$. We see from Fig.\ \ref{fig:eveccoherence2} (a) that in the Fock regime we have a ground state of the form $\vert \Psi^{0} \rangle \approx (\vert 0,55\rangle+\vert 1,45\rangle)/\sqrt{2}$, where two $\delta$-function-like spikes are separated by a distance $\Delta N_{R}=10$ in Fock space. Although perhaps not a fully fledged Schr\"{o}dinger cat state, it is a step in that direction.

If instead one sets $W=100 \, U$, then one could in theory achieve the  ``NOON'' state \cite{mitchell04}  $\vert \Psi^{0} \rangle \approx (\vert 0,100\rangle+\vert 1,0\rangle)/\sqrt{2}$. However, such a state would be extremely delicate and would quickly succumb to decoherence. Interestingly, such a state is also difficult to compute because its energy difference from the first excited state is exponentially small and numerical diagonalization routines find them hard to distinguish. This situation is very similar to that found when computing macroscopic self-trapping in double-well systems (no impurity) \cite{krahn09,odell01}, where split number distributions are also formed. The problem can be diagnosed by checking if the parity of the eigenstates is maintained: the ground state should have even parity in the impurity location and the first excited state odd parity in the impurity location, like in Fig.\ \ref{fig:evecsig0}, due to the symmetry of the hamiltonian. Numerical errors can cause this parity to be lost when the states become nearly degenerate. One remedy for this situation is to form even and odd combinations of the (erroneously) computed ground and first excited states, so that parity is enforced. This method was applied for the smaller values of $J^{a}$ shown in Figs.\ \ref{fig:coherence3} and \ref{fig:eveccoherence2}.

\begin{figure}
\includegraphics[width=0.9\columnwidth]{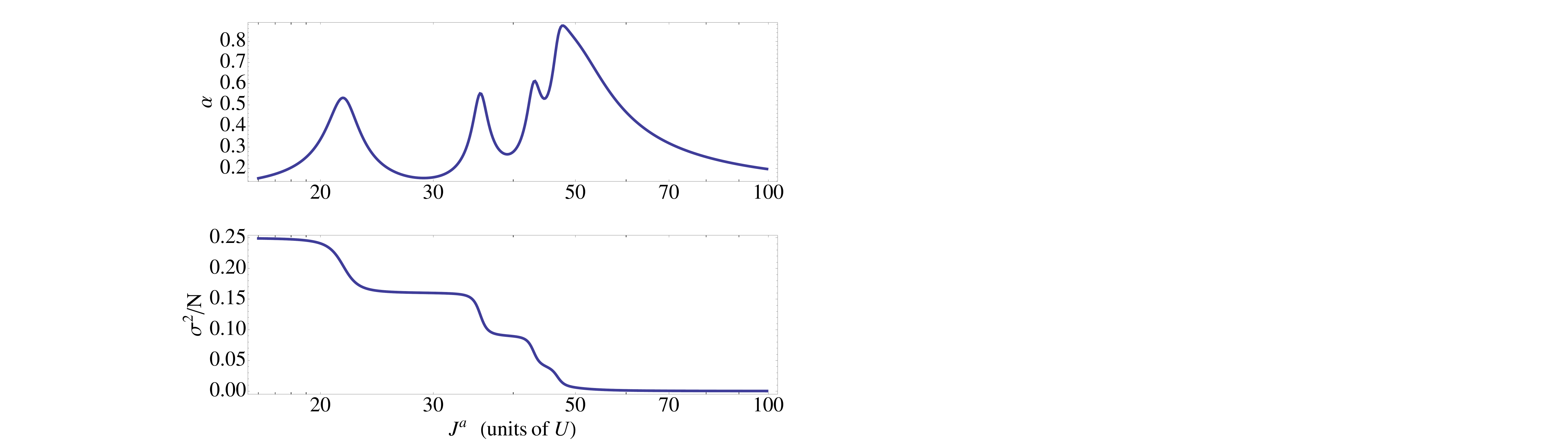}
\caption{(Color online) The coherence $\alpha$, and the square of the number fluctuations $\sigma^2$, in the ground state as a function of the impurity hopping energy $J^{a}$. In this figure $N=100$, $J/U=0.001$, $W/U=10 $, and the tilt is set to zero. This plot is a zoom-in of Fig.\ \ref{fig:coherence3}.}
	\label{fig:coherence4}
\end{figure}

The coherence and number fluctuations when $W = 10 \, U$ are shown in Fig.\ \ref{fig:coherence3}. Four peaks can be distinguished in the coherence, although the last one is quite wide and contains several features (a zoom-in of Fig.\ \ref{fig:coherence3} is given in Fig.\ \ref{fig:coherence4}). As before, the peaks in the coherence  coincide with steps in the number fluctuations, and the three panels of Fig.\ \ref{fig:eveccoherence2} show the many-body wave function in three different regimes. Panel (b) corresponds to the maximum coherence, which is $\alpha=0.87$, and is located at $J^{a}=48$. The wave function at this point is remarkably wide considering the very small value of $J = 0.001 \, U$. The mean-field prediction for the bifurcation point is $J^{a}=50 \, U$, which, once again, gives a good estimate for the parameter values at which dramatic changes in the many-body wave function occur.

\section{Entanglement entropy}

One measure of the degree of entanglement between two subsystems is provided by their entanglement entropy. This can be calculated from the reduced density matrix for either subsystem. The reduced density matrix for the bosons is defined to be
\begin{equation}
	\hat \rho_B  \equiv \sum\limits_{M_{R}=0,1} \langle M_{R} |\hat\rho | M_{R}\rangle = \sum\limits_{M_{R}=0,1} \langle M_{R}|\psi\rangle\langle\psi |M_{R}\rangle
\end{equation}
where
\begin{equation}
\vert \Psi \rangle = \sum_{M_{R},N_{R}} C_{M_{R},N_{R}} \vert M_{R}, N_{R} \rangle
\end{equation}
is the many-body wave function in the joint number basis $\vert M_{R}, N_{R} \rangle \equiv \vert M_{R} \rangle \bigotimes \vert N_{R} \rangle$. One finds that
\begin{equation}
	\hat \rho_B = \sum\limits_{N_{R},N_{R}^\prime}\left(C_{0,N_{R}^{\prime}}^* C_{0,N_{R}} + C_{1,N_{R}^{\prime}}^* C_{1,N_{R}}\right) |N_{R}\rangle\langle N_{R}^\prime| \, .
\end{equation}
The matrix elements of the reduced density matrix for the bosons in the number basis are thus
\begin{equation}
	\langle N_{R} |\hat\rho_B|N_{R}^\prime\rangle = C_{0,N_{R}^{\prime}}^*C_{0,N_{R}} + C_{1,N_{R}^{\prime}}^*C_{1,N_{R}} \, .
\end{equation}
The entanglement entropy $S_B$ is defined in terms of the reduced density matrix as
\begin{equation}
	S_B = -\sum\limits_{N_{R}} \langle N_{R}|\hat\rho_B \ln \hat\rho_B |N_{R}\rangle = -\sum\limits_{i} \rho_{B_i}\ln \rho_{B_i}
\end{equation}
where $\rho_{B_i}$ are the eigenvalues of $\hat \rho_B$. The entanglement entropy of the bosons is the same as that of the equivalent quantity for the impurity
\begin{equation}
S_I = -\sum\limits_{i} \rho_{I_i}\ln \rho_{I_i}
\end{equation}
where the matrix elements of the reduced density matrix for the impurity are
\begin{equation}
	\langle M_{R}|\hat\rho_I|M_{R}^{\prime}\rangle = \sum_{N_{R}} C_{M_{R}^{\prime},N_{R}}^* C_{M_{R},N_{R}} \, .
\end{equation}
Because of the equality of $S_{I}$ and $S_{B}$, we shall refer to the entanglement entropy simply as S rather than $S_B$ or $S_I$.

\begin{figure}
		\includegraphics[width=0.9\columnwidth]{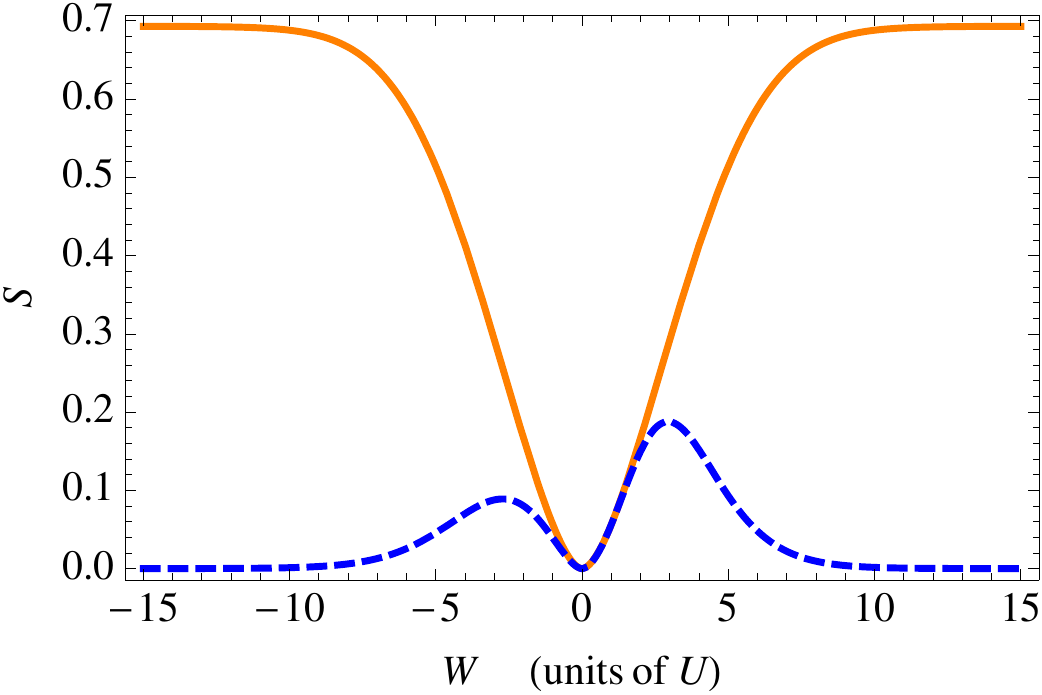}
	\caption{(Color online) Entanglement entropy $S$ as a function of $W$ for $N=100$ and $J=J^{a}=1.5 \, U$.  The solid orange curve is for zero tilt $\Delta \epsilon=\Delta \epsilon^{a} = 0$, and the dashed blue curve is for finite tilt $\Delta \epsilon=\Delta \epsilon^{a} = U$.}
\label{fig:entropy}
\end{figure}

In Fig.\ \ref{fig:entropy} we plot the entanglement entropy of the ground state as a function of $W$ for $N=100$. The solid orange curve is for a perfectly balanced double well and the dashed blue curve is for a tilted double well with $\Delta \epsilon=\Delta \epsilon^{a} = U$. In both cases the entropy is zero at $W=0$, because there is no entanglement when the boson-impurity interaction vanishes, but the two curves display very different behavior for large $\vert W \vert $. For the balanced case the entanglement entropy approaches $S=\ln 2$ as $W \rightarrow \pm \infty$ because in that regime the bosonic ground state probability distribution in Fock space tends to two $\delta$-function peaks at $N_{R}=0$ and $N_{R}=100$ for both the repulsive and attractive cases. Therefore, the system can occupy two states each having a probability of 0.5 and the entropy becomes $S=-2 \times 0.5 \ln 0.5 = \ln 2$. In the tilted case the ground state tends to a single $\delta$-function peak in bosonic Fock space as $W \rightarrow \pm \infty$ and therefore the entanglement entropy tends to zero in those limits. Of course, the tilt can never be precisely zero in an experiment and so in reality we always expect $S \rightarrow 0$ if the limit $W \rightarrow \pm \infty$ can be achieved. However, as the tilt becomes small the value of $\vert W \vert$ required for the system to become sensitive to a finite value of $\Delta \epsilon$ or $\Delta \epsilon^{a}$ becomes large.

Once again, it is interesting to ask whether the mean-field prediction for the bifurcation point given by Eq.\ (\ref{eq:loopcondition}) bears any relevance to the many-body results. For the parameters used in Fig.\ \ref{fig:entropy}, we find that $W_{c}= \pm 1.76 \, U$ (this result is only for the zero tilt case). These values do not correspond to any obvious features on the solid orange curve in Fig.\ \ref{fig:entropy}, but we note that the first derivative of this curve with respect to $W$ has extrema close by at $W= \pm 2.68 \, U$.

\section{Summary and Discussion}
\label{sec:conclusions}

In this paper we have studied the effects a single impurity can have upon a BEC in a double well potential, with the emphasis  placed upon reconciling the many-body and mean-field descriptions. The mean-field theory is nonlinear (indeed, our system is analogous to a double pendulum, which is known to be chaotic) and the static solutions undergo bifurcations  as parameters such as the tilt $\Delta \epsilon$, and the boson-impurity interaction $W$, are varied. 
In particular, the lowest lying mean-field solution  (corresponding to both pendula pointing down) undergoes a pitchfork bifurcation when $W$ is increased past a critical value $W_{c}$ given by Eq.\ (\ref{eq:loopcondition}). 
This bifurcation is due to the spontaneous formation of an imbalance in the number of atoms in the two wells. Considered as a function of the tilt between the two wells, the bifurcated solutions form a swallowtail loop in the lowest lying mean-field solution. This critical value of the boson-impurity interaction need not be large: taking numbers from the experiments \cite{albiez05} and \cite{levy07} and extrapolating them to include an impurity allows one to estimate that $W_{c} \approx 10 \ U$.

 The bifurcation in the lowest lying mean-field solution corresponds, in the many-body theory, to a splitting of the ground state atom number probability distribution into two separate pieces. One piece corresponds to the impurity being in the left well  and an increased (for positive values of $W$) or decreased (for negative values of $W$) number of bosons in the right well, and vice versa for the other piece. Unlike the mean-field solution, the many-body solution is in general a superposition of both cases (at least until decoherence is added into the model), and in the large $W$ limit takes the form of a Schr\"{o}dinger cat state.

The presence of the impurity can have a dramatic effect on the coherence $\alpha$ of the bosons between the two wells: at the bifurcation $\alpha$ is strongly peaked, even for exceedingly small values of the bosonic hopping energy $J$. This phenomenon provides a readily verifiable experimental signature of the presence of the bifurcation. A single particle tunneling between the two wells can therefore drive the system into a coherent state \cite{bausmerth07,zapata03}. However, either side of the bifurcation $\alpha$ returns to the very low background level set by $J$, and, surprisingly, this is true even when $J^{a}$ is very large so that the impurity is highly mobile. It is conceivable that the high coherence at the bifurcation in a BEC-impurity system could be usefully employed in interferometry \cite{baumgartner10}, where one wants both high coherence and yet also small number fluctuations (because the boson-boson interactions mean that different boson numbers in the wells lead to different mean-field shifts).
 
From the quantum measurement theory perspective, the formation of a Schr\"{o}dinger cat state is necessary if the BEC is to act as a macroscopic device that measures the position of the microscopic impurity. For, only then can the BEC unambiguously distinguish between the impurity being in the left and right wells (the Schr\"{o}dinger cat state is assumed to collapse onto one of its two composite states under the influence of an environment).  Given that the bifurcation heralds the splitting of the number distribution, it is interesting to inquire about the fate of the bifurcation in the  macroscopic (large $N$) limit. Examining Eq.\ (\ref{eq:loopcondition}), we see that if $N U \gg J$ , then $W_{c}$ tends to the value
\begin{equation}
\lim_{N \to +\infty} W_{c}(N) = \sqrt{2J^{a}U} \label{eq:loopconditionNinfinite}
\end{equation}
which is independent of $N$. This expression bears a resemblance to Eq.\ (\ref{eq:plasmafrequency}) for the plasma frequency, which gives the energy of the first excited state of a BEC in a double well in the same limit, except that $J$ has been replaced by $J^{a}$ and $N$ has been set to unity. 

Taken at face value, Eq.\ (\ref{eq:loopconditionNinfinite}) seems to imply the nonsensical result that a single impurity with a finite interaction $W$ can have a finite effect on an infinitely large system. However, this interpretation is misleading. In particular, Eq.\ (\ref{eq:loopconditionNinfinite}) is written in terms of microscopic parameters and should instead be expressed in terms of intensive quantities that have meaning in the thermodynamic limit $N \rightarrow \infty$, $V \rightarrow \infty$, but $N/V=$constant. To accomplish this we note that $W$ is the interaction energy per particle, and should properly be compared with the plasma energy per particle $\hbar \omega_{\mathrm{plas}}/N$. This gives
\begin{equation}
\lim_{N \to +\infty} W_{c}(N) = \frac{\hbar \omega_{\mathrm{plas}}}{N}\sqrt{N} \, .
\end{equation}
where we have put $J=J^{a}$. Taking the plasma energy per particle as an intensive quantity that is independent of $N$, we find that $W_{c}$ scales as $\sqrt{N}$ as $N \rightarrow \infty$, and thus an infinite boson-impurity interaction is required to trigger a bifurcation in the thermodynamic limit.

\begin{acknowledgments}
We gratefully acknowledge discussions with M.K. Oberthaler, B. Prasanna Venkatesh, D. Thompson, and J.H. Thywissen. Funding was provided by the Natural Sciences and Engineering Research Council of Canada (NSERC)
and by the German Academic Exchange Service (DAAD).
\end{acknowledgments}

\appendix
\section{The double pendulum}
\label{sec:doublependulum}
For the convenience of the reader, in this Appendix we summarize some results concerning the double pendulum, which is a system made up of one pendulum suspended from another. Each pendulum consists of a massless rod
 of length $l_{i}$, a bob of mass $m_{i}$, and subtends an angle $\theta_{i}$ to the downward vertical. The upper pendulum corresponds to $i=1$ and the lower pendulum to $i=2$. The kinetic $T$ and potential $V$ energies for this system are
\cite{kibble85}
\begin{equation}
T =  \frac{m_{1} l_{1}^{2} \dot{\theta}_{1}^{2}}{2}+\frac{m_{2}}{2}\left[l_{1}^{2}\dot{\theta}_{1}^{2}+ l_{2}^{2}\dot{\theta}_{2}^{2}+2l_{1}l_{2} \dot{\theta}_{1} \dot{\theta}_{2} \cos (\theta_{1}-\theta_{2}) \right]  
\end{equation}
\begin{equation}
V  = - (m_{1}+m_{2})gl_{1} \cos \theta_{1}-m_{2}g l_{2} \cos \theta_{2} \, .
\end{equation}
In the main part of the text, we describe the BEC-impurity system in terms of the phase angles $\alpha$ and $\beta$, and their conjugate number differences $Y$ and $Z$. In order to express the double pendulum in terms of conjugate variables, we form the lagrangian $L=T-V$, and obtain the conjugate momenta via $p_{i}=\partial L / \partial \dot{\theta}_{i}$. We find
\begin{eqnarray}
p_{1} & = & (m_{1}+m_{2})l_{1}^{2} \dot{\theta}_{1}+m_{2}l_{1}l_{2} \dot{\theta}_{2} \cos (\theta_{1}-\theta_{2}) \\
p_{2} & = & m_{2} l_{2}^{2} \dot{\theta}_{2}+m_{2}l_{1}l_{2} \dot{\theta}_{1} \cos (\theta_{1}-\theta_{2}) \, .
\end{eqnarray}
Solving for $\dot{\theta}_{1}$ and $\dot{\theta}_{2}$, we can eliminate these angular velocities from the hamiltonian $H=\dot{\theta}_{i} p_{i} - L$ in favor of the conjugate momenta to give
\begin{eqnarray}
H_{\mathrm{dp}} & = & \frac{l_{1}^{2}p_{2}^{2}(m_{1}+m_{2})+l_{2}^{2}p_{1}^{2}m_{2}-2l_{1}l_{2}p_{1}p_{2} m_{2} \cos(\theta_{1}-\theta_{2})}{2 l_{1}^{2}l_{2}^{2}m_{2}[m_{1}+m_{2} \sin^{2} (\theta_{1}-\theta_{2})]} \nonumber \\
 & &- (m_{1}+m_{2})gl_{1} \cos \theta_{1}-m_{2}g l_{2} \cos \theta_{2} \, .
\end{eqnarray}
We have some freedom to choose the two masses since, when uncoupled, the frequencies of the two pendula do not depend upon them. In particular, when $m_{1} \gg m_{2} $ we have
\begin{eqnarray}
H_{\mathrm{dp}} & \approx & \frac{1}{2} \left\{ \frac{p_{1}^{2}}{m_{1} l_{1}^2 }+\frac{p_{2}^{2}}{m_{2} l_{2}^{2}}-\frac{2}{m_{1}}\frac{p_{1}}{l_{1}}\frac{p_{2}}{l_{2}} \cos(\theta_{1}-\theta_{2}) \right\} \nonumber \\
 & & - m_{1}gl_{1} \cos \theta_{1}-m_{2}g l_{2} \cos \theta_{2} \, . \label{eq:dp2}
\end{eqnarray}
For example, we might take the more massive pendulum to correspond to the BEC and the less massive one to the impurity. Providing the two pendula are close to stationary points, i.e.\ where both of them are either pointing downwards or upwards or one is pointing upwards and the other downwards, the $\cos(\theta_{1}-\theta_{2})$ term can be replaced with $\pm 1$, as appropriate. This allows for a close correspondence with the mean-field hamiltonian (\ref{eq:twomodehamiltonian}), although the sign of the boson-impurity interaction $W$ is then set depending upon whether $\cos(\theta_{1}-\theta_{2})$ equals $+1$ or $-1$. 

Although it might appear that the term corresponding to $p_{2}^{2}$, i.e.\ a term in $Y^2$, is missing from the  hamiltonian (\ref{eq:mfahamiltonian}), this is not the case providing $Y \ll 1$. For then we can expand the $\sqrt{1-4 Y^2}$ term to give
\begin{eqnarray}
H & \approx & UZ^{2} - J  \sqrt{N^2-4Z^{2}} \cos{\beta} + 2J^{a}Y^2  -J^a  \cos{\alpha} \nonumber  \\
	&&  + 2 W Y Z  \, . \label{eq:mfahamiltonianappendix}
\end{eqnarray}
where we have put $\Delta\epsilon= \Delta\epsilon^a=0$. Furthermore, putting $\sqrt{N^2-4Z^2} \rightarrow N$, which assumes $NU \gg J$, we obtain a hamiltonian equivalent to Eq. (\ref{eq:dp2}).

\end{document}